\documentclass[pre,showpacs,floatfix,nofootinbib,twocolumn]{revtex4-1}

\usepackage{graphicx}
\usepackage{hyperref}
\usepackage{bm}
\usepackage[applemac]{inputenc}
\usepackage{amsmath}
\usepackage{amssymb}
\usepackage{latexsym}
\usepackage{multirow}
\usepackage{color}

\definecolor{dockerblue}{rgb}{0.11,0.56,0.98}
\definecolor{oneblue}{rgb}{0,0,0.75}

\begin{document}
\title{Geometry controlled dispersion in periodic corrugated channels}

\author{M. Mangeat}
\author{T. Gu\'erin}
\author{D.~S. Dean}

\affiliation{Laboratoire Ondes et Mati\`ere d'Aquitaine (LOMA), CNRS, UMR 5798 / Universit\'e de  Bordeaux, F-33400 Talence, France}

\date{\today}

\bibliographystyle{apsrev}

\begin{abstract}
The effective diffusivity $D_e$ of tracer particles diffusing in periodically corrugated axisymmetric two and three dimensional channels is studied. The majority of previous studies of this class of problems are based on perturbative analyses about narrow channels, where the problem can be reduced to an effectively one dimensional one. Here we show how to analyze this class of problems using a much more general approach which even includes the limit of infinitely wide channels. Using the narrow and wide channel asymptotics, we provide a Pad{\'e} approximant scheme that is able to describe the dispersion properties of a wide class of channels. 
Furthermore, we systematically identify all the exact asymptotic scaling regimes of $D_e$ and the accompanying physical mechanisms that control dispersion, clarifying the distinction between smooth channels and compartmentalized ones, and identifying the regimes in which $D_e$ can be linked to first passage problems. 
\end{abstract}
 

\maketitle
\section{Introduction} How fast does a cloud of tracer particles, moving stochastically in a complex heterogeneous medium, disperse ? This question naturally appears in a wide range of contexts, including mixing 
 \cite{leBorgne2013stretching,dentz2011mixing,barros2012flow}, sorting \cite{bernate2012stochastic}, contaminant spreading \cite{brusseau1994transport} or chemical reactions kinetics \cite{Condamin2007}. The characterization of dispersion properties, which result from a non-trivial interplay between the geometry of the heterogeneous medium and the transport by forces and/or flows, is an active field of research \cite{leBorgne2013stretching,dentz2011mixing,barros2012flow,leitmann2017time,bernate2012stochastic,aminian2016boundaries,haynes2014dispersion1,tzella2016dispersion,guerin2015}.
At large  length and time scales, dispersion is usually characterized by an effective  diffusion tensor whose components can be considerably different from typical microscopic diffusivities \cite{brenner2013macrotransport}; canonical examples for increased and decreased diffusivities are given by, respectively, motion  in shear hydrodynamic flows (called Taylor dispersion \cite{taylor1953dispersion}) and in periodic \cite{VanKampen1992} and random \cite{dean2007effective} potentials. 

Here, we consider diffusion of non-interacting particles in channels of non-uniform cross-section, a paradigm for diffusion in confined environments \cite{burada2009diffusion,malgaretti2013entropic}, arising in contexts as varied as biological cells \cite{bressloff2013stochastic,holcman2013control}, zeolites, porous media, ion channels and microfluidic devices. It is well known that, in the absence of hydrodynamic flow, the effective diffusivity of particles in channels is lower than the microscopic diffusivity. Qualitatively, this can be understood by considering the entropy $S(z)$, which measures the number of available lateral configurations at fixed longitudinal position $z$: the narrow regions have a reduced entropy and act as entropic barriers, while the wide regions can be viewed as entropic traps, leading to a motion slower than in a uniform channel. 

The first quantitative results on diffusion in channels are attributed to Jacobs \cite{jac1967} who derived the first form of the so-called Fick-Jacobs (FJ) approximation. This standard approach, and its various extensions \cite{zwanzig1992diffusion,reguera2001kinetic,kalinay2006corrections,kalinay2005extended,kalinay2005projection,kalinay2010mapping,martens2011entropic,bradley2009diffusion,berezhkovskii2011time,dagdug2012projection,valdes2014fick}, 
are based on a dimensional reduction, and approximate the dynamics of the tracer longitudinal position $z(t)$ by a diffusive dynamics in an entropic potential $\phi(z)\equiv-TS(z)$, possibly with a position dependent diffusion coefficient $D(z)$. Once the dimensional reduction is carried out, the effective diffusion constant can be computed using exact one-dimensional results \cite{lifson1962self,reimann2001giant,reguera2006entropic}. 
Such FJ-like approaches  rely however on the assumption that the equilibration dynamics of the lateral position is fast compared to longitudinal motion, which unavoidably leads to a limited range of validity.
It has been recognized that the case of abrupt changes of channel radius requires an improvement the one-dimensional description  at the cost of employing more sophisticated methods \cite{berezhkovskii2009one,antipov2013effective,kalinay2010mapping}. 
A different picture, in principle valid for channels constituted of pores separated by narrow necks, relies on the assumption that the motion is controlled by the first passage events of tracer particles between pores.  
Calculations of effective diffusivities that rely on first passage time (FPT) arguments have been so far restricted to particular simplified geometries, such as sinusoidal channels \cite{bosi2012analytical}, septate channels (made of perfectly cylindrical connected cavities) \cite{borromeo2010particle,marchesoni2010mobility}, or channels formed by overlapping circles \cite{pineda2011diffusion} or spheres \cite{berezhkovskii2003diffusivity}. 
In general however, the regimes of validity of FJ-like approximations and FPT-approaches are different, and it is therefore difficult to describe the transition between these regimes (except for the sinusoidal channel  \cite{bosi2012analytical}).  

In the present paper, we revisit theoretically the problem of dispersion in two and three dimensional axisymmetric channels of arbitrary shape. Our approach uses an exact formula of the effective diffusivity, expressed in terms of an auxiliary function that satisfies a set of partial differential equation at the scale of a single period, which we analyze using singular perturbation analysis and conformal mapping techniques. 
We systematically identify all the (exact) asymptotic scaling regimes of diffusivity and the accompanying physical mechanisms that control dispersion. In many cases, especially the case of highly corrugated channels, the dispersion coefficient is found to depend on only a few quantities related to the channel geometry rather than on the full details of its  shape. We show how the identification of regimes far outside the validity of the one-dimensional effective description can lead to an accurate description of the effective diffusivity for a wide range of parameters, via a Pad{\'e} approximant. We identify the regimes in which $D_e$ is linked to FPT problems. We also show that, depending on the behavior of the radius near the neck, we can classify channels into smoothly and highly corrugated ones, for which the effective diffusivity displays qualitatively different behaviors. 


\section{Channel geometry and general equations for the effective diffusivity}

\begin{figure}
   \includegraphics[width=8cm,clip]{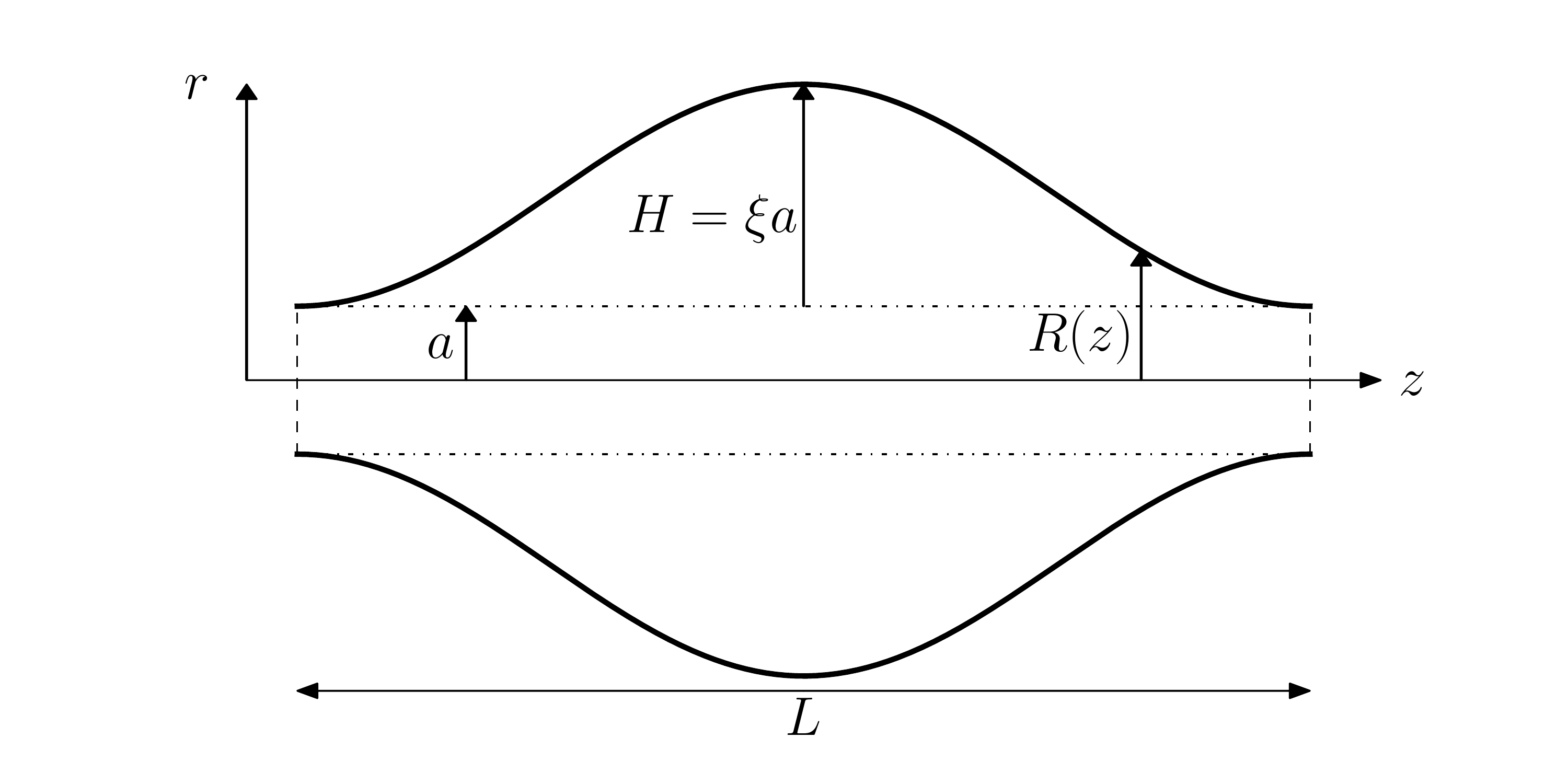}
   \caption{Schematic of a two dimensional channel of local width $2R(z)$, or slice of an axisymmetric channel in three dimensions of local radius $R(z)$. The complete channel is formed by the periodic repetition of this motif.}
   \label{schema}
\end{figure}

 We consider here the problem of the diffusion of an overdamped particle, of microscopic diffusivity $D_0$, in a two or three dimensional axisymmetric channel (Fig.~\ref{schema}), assumed to be periodic with period $L$. We denote by $z$ the (longitudinal) position in the direction parallel to the channel axis, and we assume that the channel radius $R(z)$ is parametrized as
\begin{align}
R(z)=a+ H g(z/L), 
\end{align} 
where $a$ is the minimal channel radius, $H$ is the amplitude of variation of the channel radius, and $g$ is a dimensionless periodic function of period $1$ which describes the geometrical shape of the channel boundaries, chosen to have a maximal value equal to $1$ and a  minimal value $0$. We define the dimensionless parameters, which we will see determine 
the various modes of dispersion, 
\begin{align}
\xi\equiv H/a, \hspace{0.5cm} \varepsilon\equiv a/L.
\end{align}
Channels of uniform width thus correspond to $\xi=0$, while $\xi$ is large for highly corrugated ones. 
The limit of weakly varying channels thus correspond to $\varepsilon\rightarrow0$ (at fixed $\xi$). Finally we denote by $\Omega$ the unit periodic cell, and we call $V$ its volume. 

We aim to characterize the long time effective diffusion coefficient of tracer particles  $D_e\equiv\underset{t\rightarrow\infty}{\lim}\overline{[z(t)-z(0)]^2}/(2t)$, where the overbar denotes ensemble average. 
The starting point of our analysis is the following exact expression:
\begin{align}
D_e= D_0\left(1+\frac{(d-1)\langle  f_S \ R^{d-2}\ \partial_z R\rangle}{\langle R^{d-1}\rangle}\right),\label{KuboDe}
\end{align} 
where the notation $ \langle w\rangle=\int_0^L dz w(z)/L$ is used for the uniform average over one period for any function $w$,  $d$ is the spatial dimension ($d=2$ or $3$), and $D_e$ is expressed in terms of  an auxiliary function $f_S(z) \equiv f(r=R(z),z)$, where $f(r,z)$ satisfies  
\begin{align}
&\partial_z^2f+ r^{2-d}\partial_r [ r^{d-2}\partial_r f]=0   , \label{Kubo_f}\\
&[(\partial_zR) \partial_z f-\partial_r f]_{r=R(z)}=\partial_zR,\label{BC}\\
& f(r,z+L)=f(r,z) \ ; \ \partial_rf\vert_{r=0}=0, \label{BC_Periodic}
\end{align}
where  $r$ is the distance to the central axis. These equations (\ref{KuboDe})-(\ref{BC_Periodic}) are a particular case of the general description of dispersion in arbitrary periodic systems introduced in Refs. \cite{guerin2015,Guerin2015Kubo}, they are also compatible with the equations of the macrotransport theory of Brenner and Edwards \cite{brenner2013macrotransport}. They express the macroscopic diffusion coefficient $D_e$ as a function of the microscopic structure of the channel, at the scale of one single period. 
Such a system of partial differential equations can be readily integrated numerically by using standard finite element solvers, leading to the curves presented in Figs.~\ref{sinfig} and  \ref{figinter} for various channels.  $D_e$ is represented as  a function of $\varepsilon=a/L$ for different values of the corrugation parameter $\xi$. These curves clearly display two plateaus separated by an intermediate regime; we will now study these asymptotic regimes analytically. 

\section{Slowly varying channels ($\varepsilon\rightarrow0$)} 
The first limiting case to consider is that of a slowly varying channel, which here corresponds to the limit $\varepsilon\rightarrow0$, a limit in which the FJ approximation applies since  equilibration  in the perpendicular direction is much faster than in the longitudinal direction. At leading order, a tracer particle exhibits the effectively one-dimensional dynamics of a Brownian particle $z(t)$ with diffusion coefficient $D_0$ advected by the potential $\phi(z)=-k_BT\ln(R^{d-1}(z))$. Here, the Lifson-Jackson  formula \cite{lifson1962self} provides an estimate for  the effective diffusion coefficient $D_e$:
\begin{align}
D_e \underset{\varepsilon \rightarrow 0}{=} \frac{D_0}{\langle (1+\xi g)^{d-1}\rangle\langle (1+\xi g)^{1-d}\rangle} \hspace{0.5cm} \equiv D_{\mathrm{FJ}} \label{D_FJ_SmallestOrder}
\end{align}
This well known expression clearly shows (from Jensen's inequality) that the effective diffusivity $D_e$ is reduced compared to the microscopic diffusion coefficient $D_0$, it is furthermore independent (at leading order) of the channel period $L$.  This estimate can be recovered from the equation for $f$ by a standard perturbation theory in $\varepsilon$ for $d=2$ \cite{dorfman2014assessing}, and we show in the appendix \ref{appendixA} how to generalize to $d=3$.  


\begin{figure}
	\includegraphics[width=8cm,clip]{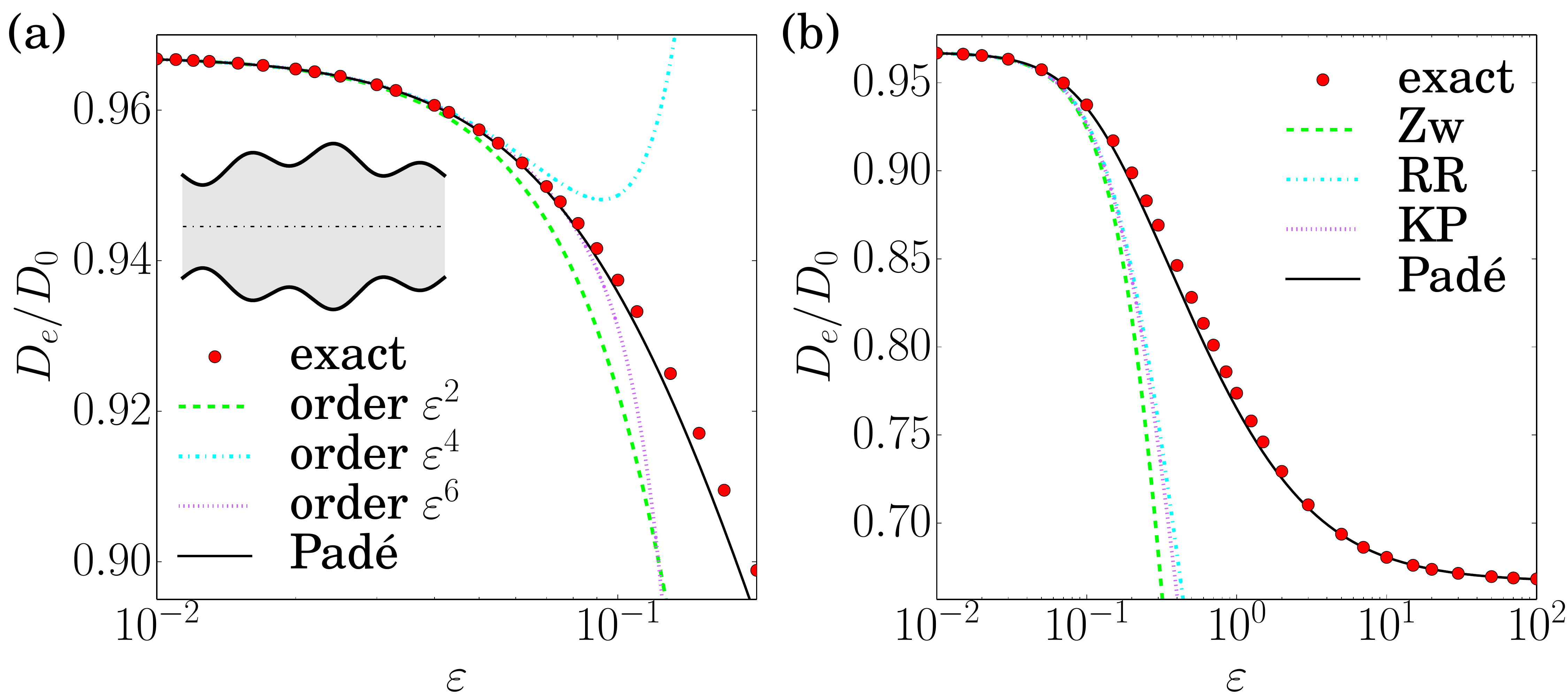}
	\caption{(color online) Effective diffusivity for the bidimensional channel of radius $R(z) =a\{1/2+0.266[\cos(2\pi z/L)+\sin(6\pi z/L)]\}$  for small (a) and finite (b) values of $\varepsilon$. The channel shape is represented in inset. On both plots,  disks represent  the numerical values of $D_e/D_0$ obtained by solving Eqs.~(\ref{KuboDe})-(\ref{BC_Periodic}), and  continuous lines correspond to the Pad\'e approximant (\ref{eqPadeapp}). In (a), the first orders of the expansion of $D_e$ in powers of $\varepsilon$, obtained from Refs.~\cite{kalinay2006corrections,dorfman2014assessing}, are represented. In (b), we also represent the results obtained by using one-dimensional re-summed formulas for the local diffusivity $D(z)$ proposed by Zwanzig (Zw) \cite{zwanzig1992diffusion}, Reguera and Rubi (RR) \cite{reguera2001kinetic} and Kalinay and Percus (KP) \cite{kalinay2006corrections}.
}\label{sinfig}
\end{figure}

\begin{figure}
	\includegraphics[width=8cm,clip]{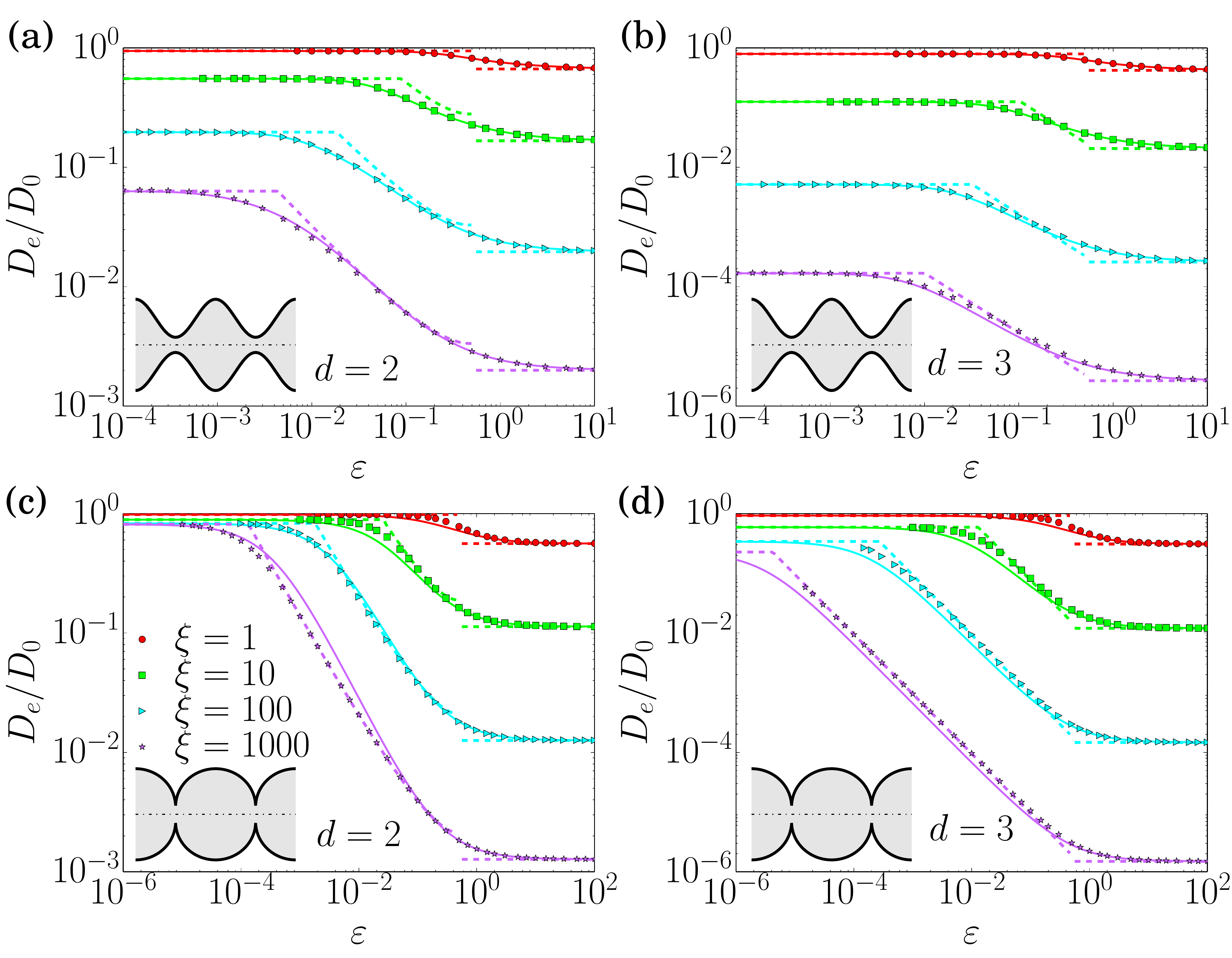}
	\caption{(color online) Effective diffusivity $D_e$ for channels of sinusoidal shape $g(u)=[1+\cos(2\pi u)]/2$ in two dimensions (a) and three dimensions (b), and ellipsoidal shape $g(u)=\sqrt{1-4u^2}$ in two dimensions (c) and  three dimensions (d). Disks represent the numerical solution of Eqs.~(\ref{KuboDe})-(\ref{BC_Periodic}), continuous lines correspond to the Pad\'e approximant (\ref{eqPadeapp}). 
Dashed lines represent the various asymptotic regimes: FJ expression  (\ref{D_FJ_SmallestOrder})  for $\varepsilon\rightarrow0$, wide channel limit (\ref{ResultLargeEps}) for $\varepsilon\rightarrow\infty$ and narrow escape regime (\ref{DeInt}) for intermediate $\epsilon$ (the value $\kappa=2/\pi$, valid for $H\gg L$, was used). 
}
	\label{figinter}
\end{figure}

Several works have attempted to improve this estimate, using various approaches. The most obvious one consists of calculating more terms in the expansion in $\varepsilon$: this has been done by assuming that the dynamics for $z(t)$ can be described  by a Markovian one, with a position dependent local diffusivity $D(z)$. Perturbation expansions  for $D(z)$ have been proposed \footnote{In Ref. \cite{kalinay2006corrections} an anisotropy of the microscopic diffusion tensor is considered, and the small parameter of the perturbation expansion is the  ratio $D_\parallel/D_\perp$;  expansions in powers of $\varepsilon$ or of this small parameter are equivalent.} which have been found to be consistent with the expansion of the macrostransport theory performed up to order $\varepsilon^4$ \cite{dorfman2014assessing}. However, such series in powers of $\varepsilon$ fail to describe the numerical curve as soon as $\varepsilon$ is not small (see Fig.~\ref{sinfig}), for the obvious reason that at large $\varepsilon$ the curve should reach a plateau instead of being polynomial. The use of Pad{\'e} approximants is a standard way to enforce a series expansion to have a constant limit at large $\varepsilon$, while retaining precision for small $\varepsilon$: it consists  of writing  $D_e =\sum_{n=1}^q a_n \varepsilon^n/\sum_{n=1}^p b_n \varepsilon^n$, with $p=q$ in order to ensure a finite limit for large $\varepsilon$, while the coefficients  $a_n,b_n$ are chosen to be consistent with the small $\varepsilon$ expansion. We have tried this procedure, but we concluded that it does not lead to accurate results, as the plateau at large $\varepsilon$ is not predicted correctly. 

Other approaches \cite{zwanzig1992diffusion,reguera2001kinetic,kalinay2006corrections} have considered different choices of $D(z)$, obtained by partial re-summation techniques, and leading to alternative estimates of $D_e$. However it is seen on Fig.~\ref{sinfig} that none of these re-summations correctly estimate $D_e$ for finite values of $\varepsilon$, and it is also known that they are not consistent with exact small $\varepsilon$ expansion \cite{dorfman2014assessing}. Therefore, FJ-like approaches are, by construction, not likely to be able to estimate $D_e$ for finite values of $\varepsilon$, which is why we focus on the  opposite limit, $\varepsilon\rightarrow\infty$ of fast varying  channels. 

\section{The limit of wide channels ($\varepsilon\rightarrow\infty$)}
 In the limit of wide channels, where $a, H\gg L$, the diffusivity at leading order can be deduced as follows. At the time scale $\tau\sim L^2/D_0$, particles at $r<a$ can be considered to diffuse freely in the longitudinal direction, while particles at $r>a$ can be considered as immobile. We can thus estimate the mean square displacement during a time $t$ to be $\overline{ z^2(t)}=2D_0T_c(t)$, where $T_c(t)$ is the time spent in the region $r<a$ up to time $t$. Ergodicity implies that $T_c(t)/t$ is also the ratio of the volume of the region $r<a$ to the total volume of the periodic cell, which leads to 
\begin{align}
D_e=D_0 \frac{a^{d-1}}{\langle  (a+H g)^{d-1}\rangle}. \label{ResultLargeEps}
\end{align}
This expression is the same as that found in comb-like geometries \cite{berezhkovskii2015biased,dean1993brownian} or tubes with dead-end regions \cite{dagdug2007transient} in simplified geometries. However, this argument should hold only for infinitely thin dead-end regions, and it does not take into account tracer particles that cross the hypersurface $r=a$, and corresponding corrections to the effective diffusivity are not easy to estimate. In what follows we carry out a  quantitative analysis of the exact equations  (\ref{Kubo_f}) in the large $\varepsilon$ limit. 

In order to construct  the auxiliary function $f$ in the limit of wide channels $\varepsilon\rightarrow\infty$, it is convenient to use rescaled variables, $\tilde{z}=z/L$ and $\tilde{r}=r/a$, in which case the variation range of variables $\tilde{z}, \tilde{r}$ is independent on $\varepsilon$. At leading order in $\varepsilon$, the resulting equation   (\ref{Kubo_f}) for $f(\tilde{r},\tilde{z})$ becomes $\partial_{\tilde{z}}^2f=0$, which, using the boundary conditions (\ref{BC}) and (\ref{BC_Periodic}) leads to solutions of the form
 \begin{align}
	f(\tilde{r},\tilde{z})= \tilde{z}\ \theta(\tilde{r}-1)L+b(\tilde r) \label{f_Far}
\end{align}
where $\theta$ is the  Heaviside function and $b(\tilde r)$ is an undetermined function of $\tilde{r}$. The above solution for $f$ is not satisfactory because it is not continuous at $\tilde{r}=1$.  This is the signal of the presence of a boundary-layer near $r=a$. The  size of the boundary layer for the lateral variable $r$ is found by inspection to be $L$: it corresponds to the region in which tracer particles can cross the line $r=a$ in the time $L^2/D_0$ needed by tracer particles to reach a neighboring pore. It is now useful  to write $r=a+\eta L$, in which case the equation  (\ref{Kubo_f}) becomes at leading order in $\varepsilon$
\begin{align}
&\partial_\eta^2f+\partial_{\tilde{z}}^2f=0\\ 
&\begin{cases}
\partial_{\tilde{z}}f\vert_{\tilde{z}=\pm1/2 }= L & (\eta>0)\\
f(\eta,\tilde{z}=1/2)=f(\eta,\tilde{z}=-1/2)  & (\eta<0)
\end{cases}\label{BCBoundaryLayer}
\end{align}
(note that  this equation holds in dimensions 2 and 3). Furthermore, to match with the outer solution (\ref{f_Far}), $f$ must behave as $f\simeq b(1)+\tilde{z} L$ for $\eta\rightarrow\infty$, and $f$ must be constant for $\eta\rightarrow-\infty$. This problem can now be handled by the use of complex analysis: we look for a solution $f=\mathrm{Re}(w(Z))$, where $w$ is an analytic function of the complex variable $Z=\tilde{z}+i\eta$. If we make the transformation $Z_1= i e^{i \pi Z}$, the problem becomes equivalent to the two-dimensional electrostatic problem consisting of finding the potential generated by two perfectly conducting neighboring horizontal plates, being located between $(\pm1,0)$ and $(\pm\infty,0)$, on which opposite values of the potential is imposed. The solution of this problem can be constructed using a Schwarz-Christoffel transform (see appendix \ref{appendixC}), and we find
\begin{align}
f=
\mathrm{Re}\left[ \frac{i L}{\pi}\ln\left(1+\sqrt{1+e^{-2\pi i \tilde{z}+2\pi\eta}}\right)\right] + b(1) ,
\end{align} 
where $\mathrm{Re}(...)$ represents the real part of a complex number. It can be checked that the above formula satisfies the boundary conditions (\ref{BCBoundaryLayer}) and matches with the outer solution (\ref{f_Far}) when one takes $\eta\rightarrow\pm\infty$. Inserting this formula into Eq.~(\ref{KuboDe}) yields 
\begin{align}
D_e \underset{\varepsilon \rightarrow \infty}{=} D_0 \frac{a^{d-1}}{\langle  (a+H g)^{d-1}\rangle}\left(1+\frac{(d-1)\ln2}{\pi \varepsilon}\right) \label{EqDeLargeEps},
\end{align}
where the $\varepsilon^{-1}$ correction comes from the contribution of $f$ in the boundary layer. These corrections, which quantify the contribution to dispersion of the particles that can cross the line separating the {\em blocked} region from the  regions of free longitudinal move, do not depend on the details of the channel geometry: they are characterized by a {\em universal} numerical constant equal to $\ln2/\pi$.

\section{An approximant including both narrow and wide channel limits}
At this stage, we can construct a Pad\'e type approximant for $D_e$,
\begin{equation}
\label{eqPadeapp}
D_e= D_{\mathrm{FJ}} \frac{1+ a_1 \varepsilon + a_2 \varepsilon^2 +a_3 \varepsilon^3}{1+ b_1 \varepsilon + b_2 \varepsilon^2 +b_3 \varepsilon^3},
\end{equation}
where the coefficients $a_i,b_i$ are carefully chosen to ensure that the expression for $D_e$ is exact for both the wide channel limit $\varepsilon\rightarrow\infty$ (up to order $\varepsilon^{-1}$, using Eq.~(\ref{EqDeLargeEps})) and the slowly varying channel limit $\varepsilon\rightarrow0$ (up to order $\varepsilon^4$, for which we used expressions in the literature \cite{dorfman2014assessing}), shown in appendix \ref{appendixB}). This approximant incorporates effects that cannot be captured by FJ-like approaches, and is found to agree with the numerical curve for almost all  values of $\varepsilon$ (see Figs.~\ref{sinfig} and  \ref{figinter}). We therefore emphasize
 that the strength of our approach is that it allows an accurate description of $D_e$ which ranges from narrow  to wide channels. 

\section{The FJ approximation for highly corrugated channels} 
We now proceed to simplifying the description of the mechanisms controlling dispersion in the limit of large ratio $\xi=H/a$ of maximum width over minimal aperture. Consider first the large $\xi$ limit of the FJ expression. 
The result of taking $\xi\rightarrow\infty$ in Eq.~(\ref{D_FJ_SmallestOrder}) depends on the existence of the integral $\int dz/g^{d-1}(z)$, which may be a divergent one (because $g$ vanishes for some value of $z$). We now assume that the behavior of $R$ near the point of minimal aperture (here taken as the origin of longitudinal axis) is characterized by 
\begin{align}
R(z\rightarrow0)\simeq a+\gamma \vert z\vert^{\nu} \label{AssumptionR}
\end{align}
where $\gamma$ is a quantity that characterizes the local geometry of  the narrowest region of the channel. For example, differentiable channel profiles correspond to $\nu=2$, in which case $\gamma$ is half the minimal curvature at the neck. If the neck is composed of  connected  conical portions (so $\nu=1$), $\mathrm{arctan}(\gamma)$ is half the opening angle of these cones. The assumption (\ref{AssumptionR}) is equivalent to 
\begin{align}
g(\tilde{z}\rightarrow0)\simeq A \vert \tilde{z}\vert^\nu \ ; \ \gamma=AH/L^\nu
\end{align}
If we define $\nu_c(d)=1/(d-1)$, we see that the integral of $1/g^{d-1}$ is infinite when $\nu>\nu_c$. 
In this case the dominant contribution in the integral $J\equiv\int_0^1d\tilde{z}/(1+\xi g(\tilde{z}))^{d-1}$ comes from the values of $z$ close from the points of smallest channel width, so that $J\simeq \int_{-\infty}^\infty d\tilde{z}/(1+\xi A \vert\tilde{z}\vert^{\nu})^{d-1}$ (where we can replace the integration bounds by $\pm \infty$ without changing the integration result at leading order). Computing this integral leads to 
\begin{align}
\frac{D_{\mathrm{FJ}}}{D_0}\simeq \frac{\nu \sin(\pi/\nu) (A\xi)^{1/\nu}}{2\pi\ \xi^{d-1}  \langle g^{d-1} \rangle} \left(\frac{\nu}{\nu-1}\right)^{d-2}, \label{DFJLargexi}
\end{align}
which can also be written as
\begin{align}
D_{\mathrm{FJ}}\simeq L^2/(2T), \label{Exp_DT}
\end{align}
with 
\begin{align}
T= \frac{V }{D_0 a^{d-1-1/\nu}\gamma^{1/\nu}}\times\frac{\pi}{2\nu\sin(\pi/\nu)} \left(\frac{2(\nu-1)}{\pi\nu}\right)^{d-2}  \label{T_NEF}.
\end{align}
We can interpret $T$ as the mean first time to reach the middle of one of the narrow regions, while the other is  reflecting. The time $T$ does not depend on the precise geometric details of the channel shape: it depends only on the volume $V$ of a single pore, on the minimal channel radius $a$ and on the parameter $\gamma$ which characterizes the geometry of the channel near the neck. In this regime, the stochastic trajectories of the tracer particles can be viewed as a continuous time random walk, where the particles spend  in each pore an average time $T/2$ which measures the rate at which the tracer particles can escape the entropic barriers formed by the narrow regions. 
Eq.~(\ref{DFJLargexi}) is known in the case $\nu=d=2$ \cite{dagdug2012diffusion,crank1979mathematics}. The mean escape time $T$ to an opening at the end of a funnel has recently been calculated  using conformal mapping techniques \cite{holcman2011narrow,holcman2012brownian,holcman2013control,guerrier2014brownian} for $\nu=2$ and $d=2,3$ and coincides with the above formula \footnote{Note that, for $d=3$ and $\nu=2$, Eq.~(\ref{DFJLargexi}) is half the result given in Refs.~\cite{holcman2013control,holcman2012brownian}. It is mentioned in Ref.~\cite{guerrier2014brownian} that a correction factor of one half should be added, but misprints in the definition of $R$ and $a$ render difficult the comparison with (\ref{DFJLargexi}}, it is interesting to see that these mean escape times are also accessible via the FJ approximation. The general formula  (\ref{T_NEF}) for $T$ for any exponent $\nu$ is new to the best of our knowledge. 

An important remark here is that $D_e$ is controlled by the time to cross the neck regions: as a consequence, Eq.~(\ref{DFJLargexi}) holds as soon as the FJ approximation is a correct description of the dynamics \textit{in the neck only rather than in the whole  channel}. The relevant longitudinal length scale $l^*$ in the neck is identified from $a\sim \gamma (l^*)^\nu$, so that $l^*\sim (a/\gamma)^{1/\nu}$; the FJ approximation is valid when $l^*\gg a$, a condition which is less constraining (for $\nu>1$) than the condition $H\ll L$ which would be required for the FJ approximation to hold in the whole channel. 

Thus, if $\nu>\nu_c$, the dispersion in the limit of slowly varying channels is controlled by the geometry at the neck. The situation is completely different in the case $\nu<\nu_c$ for which the large $\xi$ limit of $D_\mathrm{FJ}$ reads
\begin{align}
D_\mathrm{FJ}=\frac{D_0}{\langle g^{d-1}\rangle\langle g^{1-d} \rangle } \label{FJ_NonSmooth}
\end{align}
In this case, the effective diffusivity depends on the channel's geometrical shape, but not on any of the parameters $a,L,H$. This is a key difference between channels with sharp necks ($\nu<\nu_c$) or smooth necks ($\nu>\nu_c$): dispersion in sharp neck channels  is not controlled by the diffusion at the neck only. Interestingly, the case $\nu=1$ in $d=3$ dimensions is included in the regime $\nu>\nu_c$ and corresponds to a regime where the dynamics at the neck controls the transitions between pores and thus the dispersion. 

\section{Intermediate regime of dispersion}
We finally study the regime that is intermediate between the limits of small and large $\varepsilon$. It is seen on Fig.~\ref{figinter} that this intermediate regime tends to increase with increasing $\xi$, and also tends to deviate from the predictions of our Pad\'e approximant. This suggests the presence of a different mechanism that controls dispersion. We treated this case by performing a singular perturbation analysis of Eq.~(\ref{Kubo_f})-(\ref{BC_Periodic}) in the limit of small pore opening by following closely the approach of Refs.~\cite{ward1993,pillay2010asymptotic} (see appendix \ref{appendixD} for details). We obtain
\begin{align} 
D_e\simeq\frac{L^2D_0}{V}\times 
\begin{cases}
2a & (d=3)\\
\frac{\pi}{2\ln(2L\kappa/a)} & (d=2)
\end{cases}\label{DeInt} 
\end{align}
where $\kappa$ is a constant that depends on the ratio $H/L$ and on the shape of the boundary; more precisely $\ln\kappa=[R({\bf r}_0,{\bf r}_0)+R({\bf r}_1,{\bf r}_1)-2G({\bf r}_0,{\bf r}_1)]\pi/2$, 
where $G$ is the pseudo-Green's function of the domain (without opening), $R$ is the non-diverging part of this Green's function and ${\bf r}_0,{\bf r}_1$ are the positions of the openings. 
In the limit $H\gg L$, $\kappa$ reaches a constant value deduced from the Green's function in an infinite strip \cite{barton1989elements}, $\kappa=2/\pi$. 
The above formula reveals that in this intermediate regime one can again interpret the stochastic trajectories as continuous time random walks, with a dispersion coefficient satisfying the relation  (\ref{Exp_DT}), $D_e=L^2/(2T)$. In 3 dimensions, $T$ is, not surprisingly, the mean escape time through a small opening embedded in a flat plane, which does not depend on the initial position of the walker, due to the non-compact feature of space exploration by a Brownian walker in 3D \cite{Condamin2007,Benichou2008}. In 2D the situation is slightly different, because Brownian search for an opening is only marginally compact, and mean escape times depend logarithmically on the initial position \cite {Condamin2007,Benichou2008}. Comparing Eq.~(\ref{DeInt}) with recent calculations of the mean escape time in 2D domains of arbitrary shape \cite{pillay2010asymptotic} reveals that $D_e=L^2/(2T)$, where $T$ is not the global mean first passage time to a pore, but is instead the time to reach a pore, starting from the opposing opening (considered as reflecting). 
The above formula has been identified for particular geometries such as septate channels in 3D \cite{borromeo2010particle} and for channels made of overlapping spheres \cite{berezhkovskii2003diffusivity}, it has already proposed for the corresponding cases in 2D \cite{marchesoni2010mobility,pineda2011diffusion} but at leading order only. 

We obtained the formula (\ref{DeInt}) rigorously from (\ref{KuboDe}) in the limit of small pore opening in the case $\nu\le1$, but one can  see from Fig.~\ref{figinter} that is actually gives  a  good description of $D_e$ in the intermediate regime for large $\xi$ for arbitrary geometries for any channel shape, be it smooth or not. It is therefore not limited to compartmentalized channels. This can be understood by noting that the large $\xi$ limit implies that the boundaries become more and more perpendicular to the channel axis near the channel necks, and one therefore recovers the conditions of the narrow escape problem at a domain boundary. 

We end our study by drawing qualitative diagrams where the asymptotic expressions for $D_e$ are summarized, together with their validity domains. Each regime corresponds to a different physical mechanism that controls the behavior of the stochastic trajectories and thus dispersion. We stress that our approach, based on the exact expression (\ref{KuboDe}) for $D_e$, enables to obtain all the asymptotic regimes.  

\begin{figure}
	\includegraphics[width=8cm,clip]{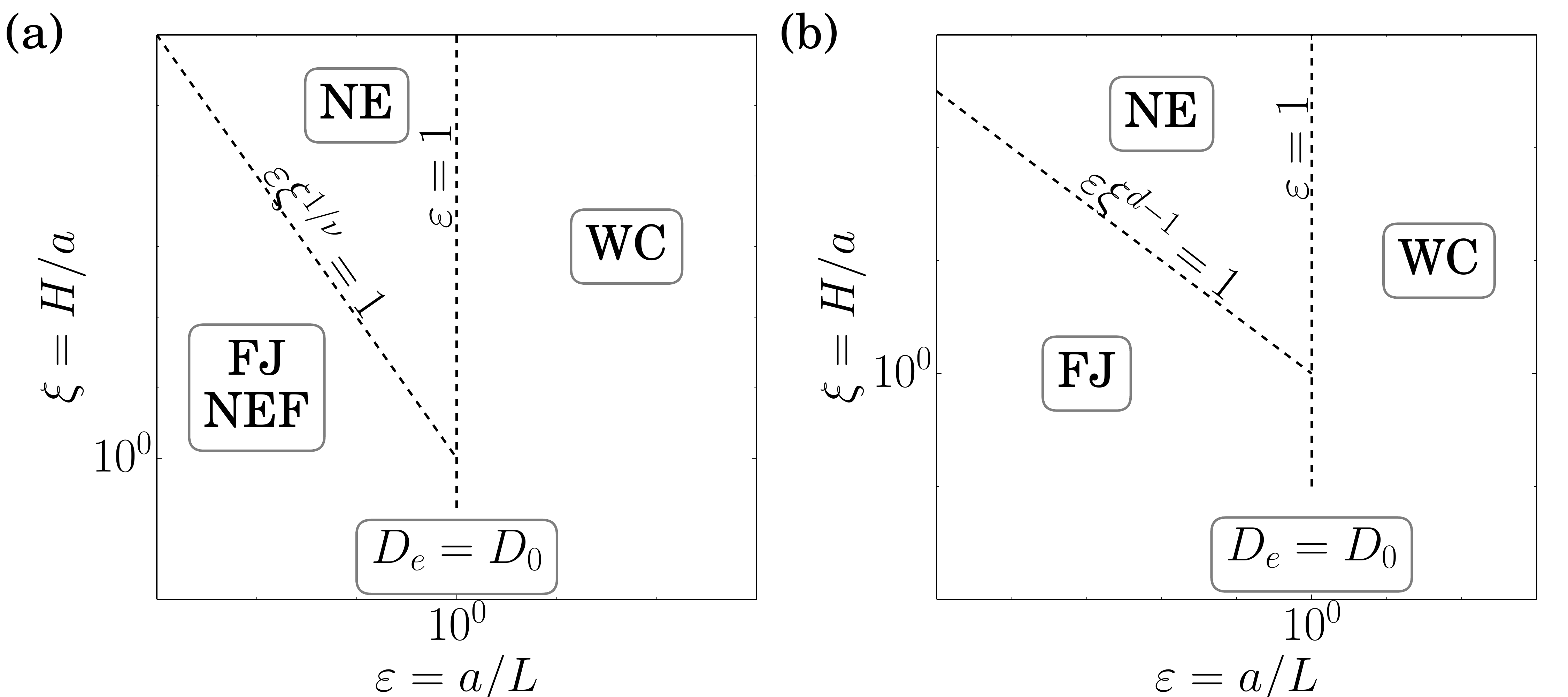}
	\caption{Diagram $(\varepsilon,\xi)$ summarizing the asymptotic estimates of $D_e$ and their validity regimes for  $\nu>\nu_c$ (a) and  $\nu \le \nu_c$ (b), with $\nu_c=1/(d-1)$. In the Fick-Jacobs (FJ), Narrow-Escape (NE) and Wide-Channel (WC) domains, the equations (\ref{D_FJ_SmallestOrder}), (\ref{DeInt}) and (\ref{ResultLargeEps}) are respectively correct. In the limit $\xi\rightarrow 0$, the effective diffusivity goes to $D_0$ for both FJ and WC limits. When $\nu>\nu_c$, the FJ expression for large $\xi$ corresponds to a regime where dispersion is controlled by events of narrow escape through a funnel (NEF). }
\end{figure}

\section{Conclusion} Let us now summarize our findings. Here we have revisited the problem of computing the effective diffusivity of tracer particles in corrugated axisymmetric two and three dimensional channels. 
We have classified the channels into two categories: smooth channels, characterized by an exponent $\nu>1/(d-1)$, for which the FJ dispersion becomes controlled by the crossing of a funnel at the necks, which we computed for any $\nu$, and non-smooth channels, with $\nu<1/(d-1)$, for which the effective diffusivity in the FJ regime becomes independent on the parameters $H,L,a$ in the strong corrugation limit. We also identified two supplementary regimes, common to all channel geometries: a comb-like regime for wide channels, where we quantified the influence on dispersion of the probability of crossing the frontier between the slow and fast regions, and an intermediate regime controlled by the standard narrow escape problem. We have also proposed a Pad{\'e} type approximant for $D_e$, which accurately describes the effective diffusivity for a wide class of parameters between the limits of narrow and wide channels. This study thus provides a refined understanding of how dispersion properties are controlled by the geometry of the channel.

\appendix
\vspace{3cm}
\begin{center}
{\bf SUPPLEMENTARY INFORMATION}
\end{center}

\section{Slowly varying channels: derivation of the FJ formula from Eqs.~(3)-(6) of the main text}
\label{appendixA}
Here we briefly describe the derivation of the effective diffusivity at leading order in the FJ regime ($\varepsilon \rightarrow 0$), starting directly from the Kubo equations for the auxilliary function $f$ [Eqs.~(3)-(6) of the main text]. We define the rescaled variables $\tilde r = r/\varepsilon$ and
 $\tilde R(z)= R(z)/\varepsilon$. The function $f$ satisfies the equation
\begin{align}
&\varepsilon^2 \partial_z^2f+ {\tilde r}^{2-d}\partial_{\tilde r} [ {\tilde r}^{d-2}\partial_{\tilde r} f]=0 \\
&[(\varepsilon^2 \partial_z \tilde R) \partial_z f-\partial_{\tilde r} f]_{\tilde r= \tilde R(z)}=\varepsilon^2 \partial_z \tilde R,\\
& f(\tilde r,z+L)=f(\tilde r,z) \ ; \ \partial_{\tilde r} f\vert_{\tilde r=0}=0.
\end{align}
We expand $f$ in powers of $\varepsilon^2$, {\em i.e.} $f(\tilde r,z)=f_0(\tilde r,z)+\varepsilon^2 f_2(\tilde r,z) +...$, and find at leading order,
\begin{align}
& {\tilde r}^{2-d}\partial_{\tilde r} [ {\tilde r}^{d-2}\partial_{\tilde r} f_0]=0 \\
&\partial_{\tilde r} f_0|_{\tilde r= \tilde R(z)}=0,\\
& f_0(\tilde r,z+L)=f_0(\tilde r,z) \ ; \ \partial_{\tilde r} f_0\vert_{\tilde r=0}=0.
\end{align}
These equations impose that $f_0$ does not depend on $\tilde r$, $f_0(\tilde r,z)=f_0(z)$. The equations at second order yield
\begin{align}
&\partial_z^2f_0+ {\tilde r}^{2-d}\partial_{\tilde r} [ {\tilde r}^{d-2}\partial_{\tilde r} f_2]=0 \label{05043}\\
&[(\partial_z \tilde R) \partial_z f_0-\partial_{\tilde r} f_2]_{\tilde r= \tilde R(z)}= \partial_z \tilde R,\label{05943}\\
& f_2(\tilde r,z+L)=f_2(\tilde r,z) \ ; \ \partial_{\tilde r} f_2\vert_{\tilde r=0}=0.
\end{align}
The solution of Eq.~(\ref{05043}) obeys $\partial_{\tilde r} f_2 = -f_0''(z) \frac{\tilde r}{d-1}$. Inserting this expression into Eq.~(\ref{05943}) yields
\begin{equation}
(d-1) \tilde R' f_0' + f_0'' \tilde R = (d-1)\tilde R'. 
\end{equation}
The solution of the above equation which takes into account the periodicity of $f_0$ obeys
\begin{equation}
f_0'(z) = 1  -\frac{\langle\tilde R^{1-d}\rangle^{-1}}{\tilde R^{d-1}}.
\end{equation}
At leading order the effective diffusivity is thus given by
\begin{align}
\frac{D_e}{D_0} &= 1+ (d-1) \frac{\langle R'(z) R(z)^{d-2} f_0(z) \rangle}{\langle R^{d-1} \rangle} +\mathcal{O}(\varepsilon^2) \nonumber \\
&= 1 - \frac{\langle \tilde R^{d-1} f_0'(z) \rangle}{\langle \tilde R^{d-1} \rangle}+\mathcal{O}(\varepsilon^2) \\
&= \frac{1}{\langle \tilde R^{d-1} \rangle\langle \tilde R^{1-d} \rangle}+\mathcal{O}(\varepsilon^2), \nonumber
\end{align}
thus recovering the Fick-Jacobs' result \cite{jac1967} for the effective diffusivity at leading order in $\varepsilon$.  

\section{Coefficients for the Pad{\'e} formula for the effective diffusivity }
\label{appendixB}

We propose in the main text to approximate the effective diffusivity by the Pad{\'e} formula
\begin{equation}
D_e= D_{\mathrm{FJ}} \frac{1+ a_1 \varepsilon + a_2 \varepsilon^2 +a_3 \varepsilon^3}{1+ b_1 \varepsilon + b_2 \varepsilon^2 +b_3 \varepsilon^3},
\end{equation}
where the coefficients $a_i,b_i$ are chosen to ensure that the expression for $D_e$ is asymptotically exact for both the narrow ($\varepsilon\to 0$) and wide channel ($\varepsilon\to \infty$) limits. More precisely, assuming that $D_e=D_\infty[1+\alpha/\varepsilon+\mathcal{O}(\varepsilon^{-2})]$ for large $\varepsilon$, and that  $D_e=D_{\mathrm{FJ}}[1+\lambda_2\varepsilon^2+\lambda_4\varepsilon^4+\mathcal{O}(\varepsilon^6)]$ for small $\varepsilon$, and denoting $K=D_{\infty}/D_{\mathrm{FJ}}$, the coefficients are
\begin{align}
a_1&=b_1=\frac{[\lambda_2^2+\lambda_4(K-1)](K-1)}{\lambda_2^2 K \alpha} \nonumber \\
a_2&=\lambda_2 - \frac{\lambda_4}{\lambda_2}\ ;  \ b_2=- \frac{\lambda_4}{\lambda_2}  \\
a_3&=Kb_3 = \frac{\lambda_2^2+\lambda_4(K-1)}{\lambda_2 \alpha} \nonumber
\label{pade}
\end{align}
The value of $D_\infty$ and $\alpha$ are identified from Eq.~(13) in the main text. The values of the $\lambda_i$ are explicitly found as follows. We consider the small $\varepsilon$ expansion of the local diffusivity $D_{1d}(z)$ in an effective one-dimensional description~\cite{kalinay2006corrections}
\begin{align}
&\frac{D_{1d}}{D_0}=\nonumber\\
& \begin{cases}
1- \frac{\varepsilon^2}{3} \tilde R'^2 + \frac{\varepsilon^4}{45}(9\tilde R'^4+\tilde R\tilde R'^2\tilde R''-\tilde R^2\tilde R'\tilde R''') & (d=2)\\
1- \frac{\varepsilon^2}{2} \tilde R'^2 + \frac{\varepsilon^4}{48}(18\tilde R'^4+3\tilde R\tilde R'^2\tilde R''-\tilde R^2\tilde R'\tilde R''')& (d=3)
\end{cases}
\end{align}
where $\tilde R=R/\varepsilon$. This formula can be inserted inserted into the Lifson - Jackson formula \cite{lifson1962self} 
\begin{equation}
\frac{D_e}{D_0}= \frac{1}{\langle \tilde R^{d-1} \rangle \langle (D_{1d} \tilde R^{d-1})^{-1}\rangle},\label{LFEq}
\end{equation}
to finally give
\begin{widetext}
\begin{align}
\frac{D_e}{D_0}= 
\begin{cases}
 \frac{1}{\langle \tilde R \rangle \langle \tilde R^{-1}\rangle} \left\{ 1-\frac{\varepsilon^2}{3} \frac{\langle \tilde R'^2 /\tilde R \rangle}{\langle \tilde R^{-1}\rangle } + \varepsilon^4 \left[ \frac{\langle \tilde R'^2 /\tilde R \rangle^2}{9 \langle \tilde R^{-1}\rangle^2 } + \frac{4 \langle \tilde R'^4/\tilde R \rangle + \langle \tilde R\tilde R''^2\rangle}{45 \langle \tilde R^{-1}\rangle} \right]+\mathcal{O}(\varepsilon^6) \right\}
& (d=2)\\
 \frac{1}{\langle \tilde R^2 \rangle \langle \tilde R^{-2}\rangle} \left\{ 1-\frac{\varepsilon^2}{2} \frac{\langle \tilde R'^2 /\tilde R^2 \rangle}{\langle \tilde R^{-2}\rangle } + \varepsilon^4 \left[ \frac{\langle \tilde R'^2 /\tilde R^2 \rangle^2}{4 \langle \tilde R^{-2}\rangle^2 } + \frac{7 \langle \tilde R'^4/\tilde R^2 \rangle + \langle \tilde R''\rangle}{48 \langle \tilde R^{-2}\rangle} \right] +\mathcal{O}(\varepsilon^6)\right\}&(d=3)
\end{cases}\label{059453}
\end{align}
\end{widetext}
from which the coefficients $\lambda_i$ can be read off. The above expansion of $D_e$ was validated for $d=2$ in Ref.~\cite{dorfman2014assessing}, and can also be found by iterating the approach of the previous section.

In the case of channels for which $R'$ can be infinite the expansion (\ref{059453}) fails as it predicts an infinite coefficient even for the $\varepsilon^2$ term. In this case, we used the following lower-order Pad{\'e} approximant, 
\begin{align}
D_e&=D_{\mathrm{FJ}}\frac{1+a_1\varepsilon}{1+b_1\varepsilon} \nonumber\\
a_1&=\frac{(D_{\mathrm{FJ}}-D_\infty)\pi}{D_{\mathrm{FJ}}(d-1)\ln2} \ ; \ b_1=\frac{(D_{\mathrm{FJ}}-D_\infty)\pi}{D_\infty(d-1)\ln2}.
\end{align}
We used this Pad{\'e} approximant for channels with elliptic boundaries in Fig.~3(c) and Fig.~3(d) in the main text. 

\section{The limit of wide channels [Derivation of Eq.~(13) in the main text]}
\label{appendixC}

\begin{figure}
	\centering
	\includegraphics[width=8cm,clip]{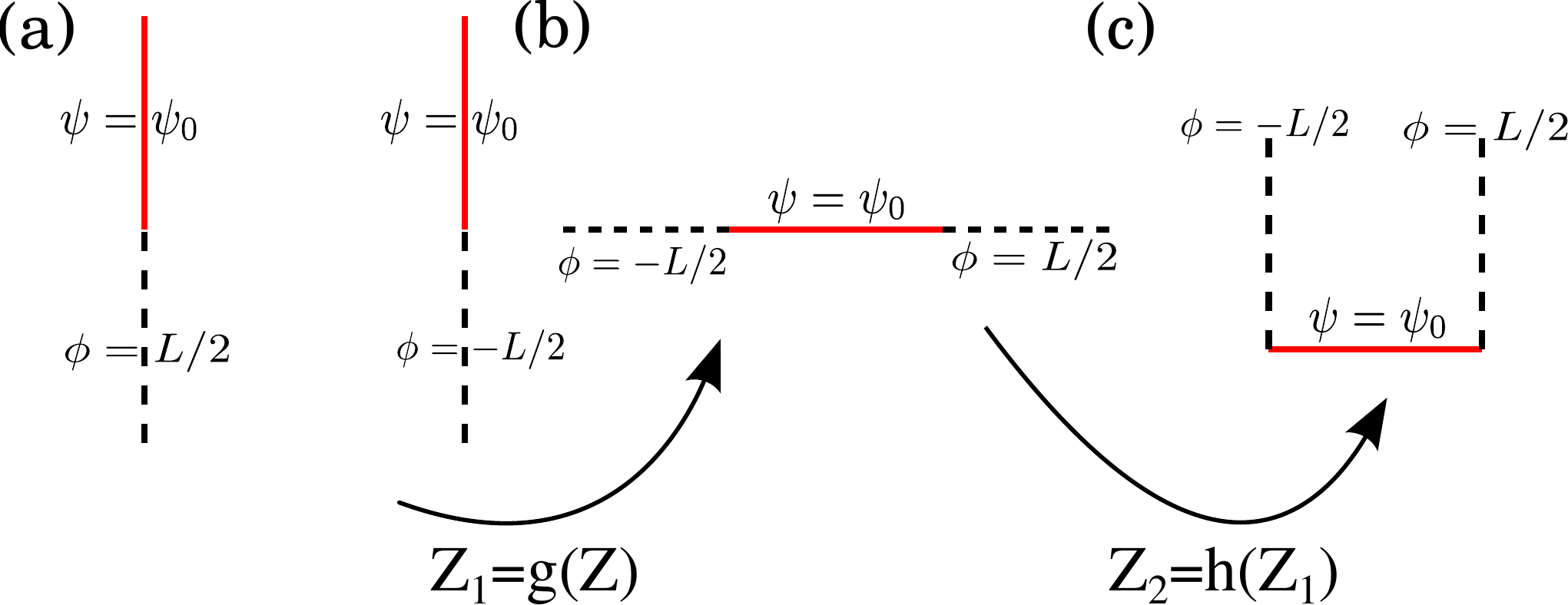}
	\caption{\label{figCM}(a) Domain in the boundary-layer, for the complex variable $Z=\tilde z+i\eta$. (b) The same domain after the conformal mapping $Z_1=g(Z)=ie^{i\pi Z}$. (c) The same domain after the Schwarz-Christoffel transformation $Z_2=h(Z_1)=\arcsin Z_1$.}
\end{figure}

In the limit of wide channels $a,H \gg L$, we can solve the Kubo formulas using complex analysis. In the boundary-layer coordinates $(\eta,\tilde z)$ defined by $r=a+\eta L$ and $z=\tilde z L$, the function $f(\eta,\tilde z)$ satisfies the Laplace equation $\partial_{\tilde z}^2 f + \partial_\eta^2 f=0$. Defining the complex variable $Z=\tilde z + i \eta$, the solution of this equation can be written as $f(\eta,\tilde z)=\mathrm{Re}( w(Z))$ where $w(Z)$ is an analytic function. Moreover, the boundary conditions $\partial_{\tilde z} f|_{\tilde z=\pm 1/2} = L$ for $\eta>0$ and $f(\eta,\tilde z=1/2)=f(\eta,\tilde z=-1/2)$ for $\eta<0$ imposes the condition on $\Phi(Z)=w(Z)-ZL-b(1)=\phi(\eta,\tilde z)+i\psi(\eta,\tilde z)$, due to the Cauchy-Riemann equations, as $\psi(\eta,\tilde z=\pm L/2)=\psi_0$ for $\eta>0$ and $\phi(\eta,\tilde z=\pm L/2)=\mp L/2$ for $\eta<0$ (see Fig.~\ref{figCM}(a)).

If we make the conformal mapping $Z_1=g(Z)=ie^{i\pi Z}$, the problem becomes equivalent to a two-dimensional electrostatic problem for two conducting horizontal plates of opposite potential, located at $\mathrm{Im} (Z_1)=0$ and $|\mathrm{Re}( Z_1)|>1$ (see Fig.~\ref{figCM}(b)). The solution of this problem can be constructed via a Schwarz-Christoffel transformation $h(Z_1)$ (see Fig.~\ref{figCM}(c)) defined by
\begin{equation}
\frac{dh}{dz}(Z_1)=\frac{A}{(Z_1-1)^{1/2}(Z_1+1)^{1/2}} = \frac{A}{\sqrt{Z_1^2-1}}.
\end{equation}
This equation can be integrated to yield $h(Z_1)=A \arcsin Z_1+B$, where $A,B$ are fixed by the relation $h(\pm 1)=\pm \pi/2$. This finally gives $h(Z_1)=\arcsin Z_1$. Defining the new coordinate in the new space as $Z_2=h(Z_1)$, we can express the solution $\Phi(Z_2)$ of the Laplace equation as a linear function of $Z_2$, where $\psi_0$ is chosen to be zero,
\begin{equation}
\Phi(Z_2)=-\frac{L}{2} + \frac{L}{\pi} \left(Z_2+\frac{\pi}{2}\right) = \frac{LZ_2}{\pi}.
\end{equation}
In terms of the original coordinates we have
\begin{equation}
\label{eqPHI}
\Phi(Z)=\frac{L}{\pi} \arcsin\left( ie^{i\pi Z}\right) = \frac{iL}{\pi} \ln\left(e^{i\pi Z} + \sqrt{1+e^{2i\pi Z}}\right),
\end{equation}
and thus the function $f(\eta,\tilde z)$ is given by
\begin{align}
f(\eta,\tilde z) &= \mathrm{Re} \left[ Z L + \frac{iL}{\pi} \ln\left(e^{i\pi Z} + \sqrt{1+e^{2i\pi Z}}\right)\right] +b(1) \nonumber \\
&= \mathrm{Re} \left[ \frac{iL}{\pi} \ln\left(1+ \sqrt{1+e^{-2i\pi \tilde z +2 \pi\eta}}\right)\right] +b(1), \label{94831}
\end{align}
which is Eq.~(12) in the main text. 

The effective diffusivity is now calculated from the expression
\begin{equation}
\frac{D_e}{D_0} = 1+(d-1) \frac{\langle R' R^{d-2} f_S \rangle}{\langle R^{d-1} \rangle}
\end{equation}
where $\langle w \rangle = \int_{-L/2}^{L/2} w(z) d z /L$, and $f_S(z)=f(R(z),z)$. This equation can also be written as
\begin{widetext}
\begin{align}
\frac{D_e}{D_0} &= 1+(d-1) \frac{\langle R' R^{d-2} [z+b(\tilde R)] \rangle}{\langle R^{d-1} \rangle}+(d-1) \frac{\langle R' R^{d-2} (f_S-z-b(\tilde R)) \rangle}{\langle R^{d-1} \rangle} \nonumber\\
&=1+ \frac{\langle (R^{d-1})' z \rangle}{\langle R^{d-1} \rangle} + \frac{(d-1)}{L \langle R^{d-1} \rangle} \int_{-L/2}^{L/2} d z R'(z) R(z)^{d-2} (f_S(z)-z-b(\tilde R)).
\end{align}
The integrand in the last integral vanishes for all $z$ that are not close to $z=\pm L/2$ (the minima of $g$), the corresponding integral can thus be calculated using the value of $f$ in the boundary layer near $z=\pm L/2$. Using the boundary-layer coordinates such that $d z R'(z) \sim d r = d \eta L$, $R(z) \sim a$ and $f_S(z)=f(\eta,z=\pm L/2)$, and performing an integration by parts on the outer integral one obtains
\begin{equation}
\frac{D_e}{D_0} =\frac{a^{d-1}}{\langle R^{d-1} \rangle} + \frac{(d-1)a^{d-2}}{\langle R^{d-1} \rangle}  \left\{ \int_{0}^{\infty} d \eta [f(\eta,z)-z-b(1)]_{z=-L/2}+\int_{\infty}^{0} d \eta [f(\eta,z)-z-b(1)]_{z=L/2} \right\}.
\end{equation}
\end{widetext}
Using Eq.~(\ref{94831}) to calculate $f(\eta,z)-z-b(1)$ in the boundary-layer, we find
\begin{equation}
\left.[f(\eta,z)-z-b(1)] \right|_{z=\pm L/2} = \mp \frac{L}{\pi} \arcsin e^{-\pi \eta},
\end{equation}
from which we finally obtain Eq.~(13) of the letter,
\begin{align}
\frac{D_e}{D_0} &=\frac{a^{d-1}}{\langle R^{d-1} \rangle} + \frac{2(d-1)a^{d-2}L}{\pi\langle R^{d-1} \rangle} \int_{0}^{\infty} d \eta \arcsin e^{-\pi \eta} \nonumber\\
&= \frac{a^{d-1}}{\langle R^{d-1} \rangle} \left[ 1 + \frac{(d-1) \ln 2}{\pi \varepsilon} \right].
\end{align}

\section{Intermediate regime of dispersion in 2D domains [Derivation of Eq.~(21) in the main text]}
\label{appendixD}

Here we identify the diffusivity in the limit of narrow openings, which turns out to be the  regime of dispersion for intermediate values of $\varepsilon$ and highly corrugated channels [Eq.~(21) in the main text]. We  follow closely the singular expansion approach of Ref.~\cite{pillay2010asymptotic} where the narrow escape problem through various openings was considered; here however we do not assume any link between the effective diffusivity and the first passage problems and we start directly from the Kubo equations for $f$, which read for a 2D channel
\begin{align}
&\partial_z^2f({\bf r}) + \partial_r^2 f({\bf r})=0 , {\bf r} \in \Omega, \label{Kubo_f_si}\\
&\vec n \cdot \vec \nabla f({\bf r}) = n_z, {\bf r} \in \partial\Omega_{out}, \label{BCout_si}\\
& f(r,z+L)=f(r,z), {\bf r} \in \partial\Omega_{in}, \label{BCin_si}\\
&\int_\Omega d r d z f({\bf r})=0,
\end{align}
where ${\bf r}=(r,z)$. $\partial \Omega$ represents the full boundary of the elementary periodically repeated pore, $\partial \Omega_{out}$ is the reflective boundary of $\partial \Omega$ and $\partial \Omega_{in}$ is the periodic boundary (corresponding to small opening) of $\partial \Omega$. If the opening does not exist ($a=0$ or equivalently $\varepsilon=0$ ),
the boundary condition (\ref{BCin_si}) no longer applies (as $\Omega_{in}=\emptyset$) and the solution is given by $f({\bf r})=z$ (up to an unimportant additive constant). For non-vanishing opening, this solution is not valid near the pore openings, since the boundary condition (\ref{BCin_si}) is no longer satisfied. We thus add a small opening pertubatively, following the approach of \cite{ward1993,pillay2010asymptotic}, and write the expansion of $f$ far from the pore openings (outer expansion) as
\begin{equation}
f({\bf r},\varepsilon) = f_0({\bf r})+ \nu_1(\varepsilon)f_1({\bf r})+\nu_2(\varepsilon)f_2({\bf r})+...,
\end{equation}
where $1 \gg \nu_i(\varepsilon) \gg \nu_{i+1}(\varepsilon)\gg \cdots $. The function $f_i({\bf r})$ ($i\ne0$) satisfies the equations
\begin{align}
&\partial_z^2f_i + \partial_r^2 f_i=0 , {\bf r} \in \Omega\\
&\vec n \cdot \vec \nabla f_i = 0, {\bf r} \in \partial\Omega_{out}\\
&\int_\Omega d {\bf r} f_i({\bf r})=0
\end{align}
Close to the openings located at ${\bf r}={\bf r}_\pm \equiv (0,\pm L/2)$, we make the change of variable $\tilde {\bf r} \equiv ({\bf r}-{\bf r}_\pm)/a$, which are both equivalent due to the periodic boundary condition (\ref{BCin_si}), and take the inner expansion of $f$ such that
\begin{equation}
f({\bf r},\varepsilon)=v(\tilde {\bf r},\varepsilon) = \mu_0(\varepsilon)v_0(\tilde {\bf r})+ \mu_1(\varepsilon)v_1(\tilde {\bf r})+\mu_2(\varepsilon)v_2(\tilde {\bf r})+...,\label{0542}
\end{equation}
and we write the matching condition
\begin{equation}
\label{matchEq}
\mu_0(\varepsilon)v_0({\bf \tilde r})+ \mu_1(\varepsilon)v_1({\bf \tilde r})+ ... \sim f_0({\bf r})+ \nu_1(\varepsilon)f_1({\bf r})+...,
\end{equation}
in the domain where ${\bf \tilde r} \cdot {\bf e_z} \rightarrow \pm \infty$ and ${\bf r} \rightarrow {\bf r}_\mp$. The function $v_i(\tilde r,\tilde z)$ satisfies the equations
\begin{align}
&\partial_{\tilde z}^2v_i({\bf \tilde r}) + \partial_{\tilde r}^2 v_i({\bf \tilde r})=0 , {\bf \tilde r} \in \tilde \Omega\\
&\partial_{\tilde z} v_i({\bf \tilde r}) = 0, {\bf \tilde r} \in \partial \tilde \Omega_{out}.
\end{align}
(note that here we assume the channel boundary to be flat near the opening, we do not consider any corrections linked to finite values of the channel curvature near the pore). In terms of  elliptic coordinates, defined as
\begin{align}
\tilde r=\cosh \mu \cos \nu\ , \ 
\tilde z=\sinh \mu \sin \nu,
\end{align}
Laplace's equation becomes $\partial_\mu^2 v_i + \partial_\nu^2 v_i=0$, with the additional boundary condition $\partial_\nu v_i=0$ on $\partial \tilde \Omega_{out}$. We now  look for solutions that are independent of $\nu$, leading to
\begin{equation}
v_i(\mu)=A_i \mu + B_i.\label{7473}
\end{equation}
where the constants $A_i,B_i$ will be identified using the matching condition. In the limit $\mu \rightarrow \pm \infty$, we obtain the behavior
\begin{align}
v_i(\tilde {\bf r}) &\underset{|{\bf \tilde r}| \rightarrow \infty , \bf{r} \rightarrow \bf{r_\mp}}{\sim} \pm A_i \ln 2 |\tilde {\bf r}| + B_i \nonumber \\
&\sim \pm A_i \ln \frac{2L}{a} \pm A_i \ln \frac{|{\bf r}-{\bf r}_\mp|}{L} + B_i.
\end{align}

From the matching condition (\ref{matchEq}), we then find $A_0=-L/2$, $B_0=0$, and
\begin{align}
&\mu_0(\varepsilon)=-[\ln (\varepsilon/2)]^{-1} = \nu_1(\varepsilon),\\
&f_1({\bf r}) \underset{{\bf r}\rightarrow {\bf r}_\pm}{\sim} \mp A_0 \ln \frac{|{\bf r}-{\bf r}_\pm|}{L}.
\end{align}
We need the expression for $f_1({\bf r})$ to find the second asymptotic term. To do this we assume that
\begin{equation}
f_1({\bf r}) \underset{{\bf r}\rightarrow {\bf r}_\pm}{\sim} \mp A_0 \ln \frac{|{\bf r}-{\bf r}_\pm|}{L} +C_\pm.
\end{equation}
From the matching condition (\ref{matchEq}), we choose $C_\pm= \mp A_1$, $B_1=0$ and
\begin{equation}
\mu_1(\varepsilon)= [\ln(\varepsilon/2)]^{-2}.
\end{equation}
Following \cite{pillay2010asymptotic}, we introduce the pseudo-Green function $G({\bf r},{\bf r}_\pm)$ defined via
\begin{align}
&\partial_z^2G + \partial_r^2 G=\frac{1}{|\Omega|}, {\bf r} \in \Omega \\
& \vec n \cdot \vec \nabla G =0, {\bf r} \in \partial \Omega_{out} \\
& G({\bf r},{\bf r}_\pm) \underset{{\bf r}\rightarrow {\bf r}_\pm}{\sim} -\frac{1}{\pi} \ln \frac{|{\bf r}-{\bf r}_\pm|}{L} + R({\bf r}_\pm,{\bf r}_\pm) \\
& \int_\Omega G({\bf r},{\bf r}_\pm) d {\bf r} =0.
\end{align}
The solution of $f_1({\bf r})$ is thus given by
\begin{equation}
f_1({\bf r}) = -\pi A_0 \left[ G({\bf r},{\bf r}_-) - G({\bf r},{\bf r}_+)  \right] + \chi.
\end{equation}
Using the behavior of $f_1$ and $G({\bf r},{\bf r}_\pm)$ close to the openings at ${\bf r}={\bf r}_-$ and ${\bf r}={\bf r}_+$, we obtain the system
\begin{align}
A_1 = -\pi A_0 \left[ R({\bf r}_-,{\bf r}_-) - G({\bf r}_-,{\bf r}_+)\right] + \chi,\\
-A_1 = \pi A_0 \left[R({\bf r}_+,{\bf r}_+) - G({\bf r}_+,{\bf r}_-) \right] + \chi.
\end{align}
and we finally find
\begin{align}
A_1 = \frac{\pi L}{4} \left[R({\bf r}_-,{\bf r}_-)+R({\bf r}_+,{\bf r}_+) - G({\bf r}_-,{\bf r}_+)-G({\bf r}_+,{\bf r}_-)\right].
\end{align}

\begin{figure}
\centering{
	\includegraphics[width=8cm,clip]{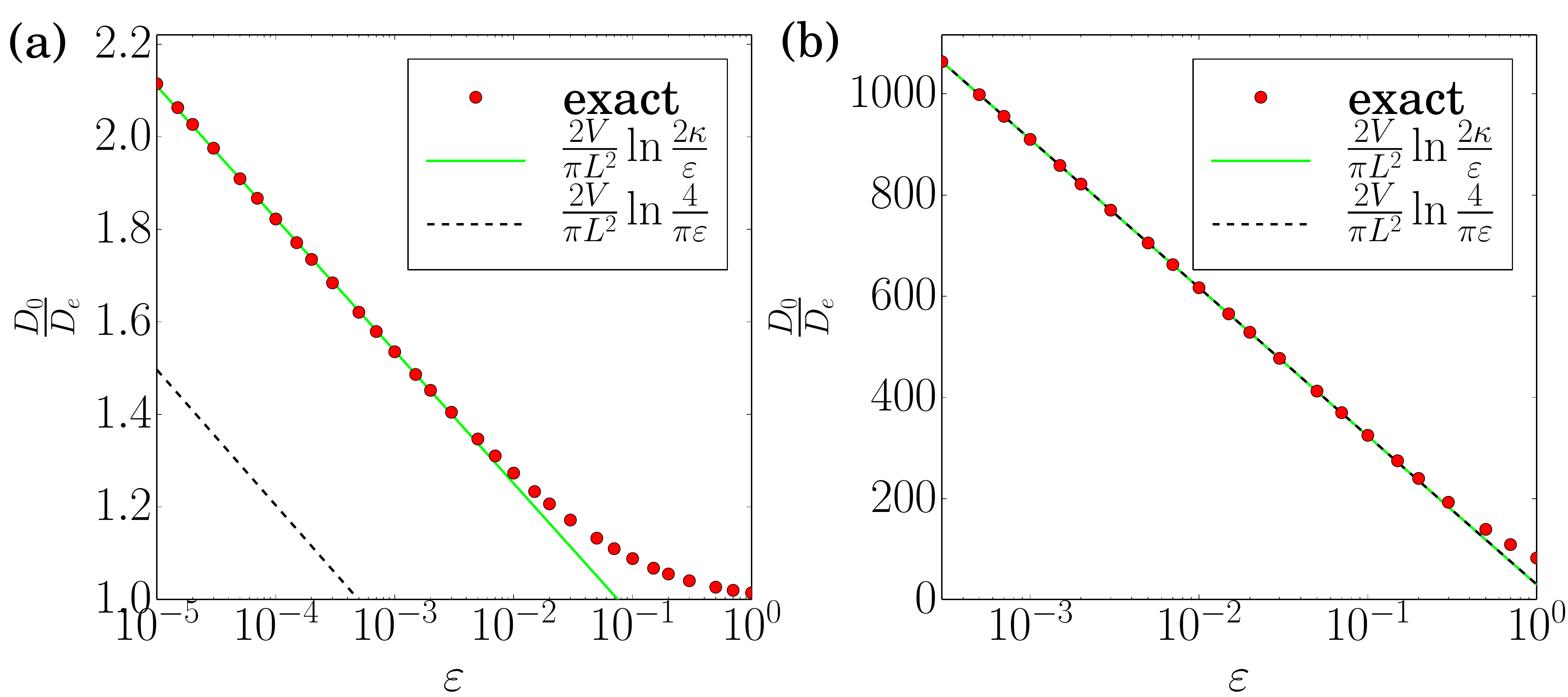}
	\caption{Behavior of $D_e/D_0$ for elliptic channels $R(z)=a+H g(x/L)$ where $g(u)=\sqrt{1-4u^2}$, $H=0.1 L$ (a) and $H=100 L$ (b) for small openings. We check  Eq.~(\ref{De_inter}). In the case $H\gg L$, the value of $\kappa$ is well approximated by the Green's function of the infinite strip: $\kappa=2/\pi$.  }}
\end{figure}

We thus have characterized the solution $f$ near the boundaries between successive pores. The effective diffusivity is, in general, given by
\begin{equation}
\frac{D_e}{D_0} = 1- \frac{1}{|\Omega|}\int_{\partial \Omega_{out}} d S_z({\bf r}) f({\bf r}).\label{058423}
\end{equation}
A few manipulations can be made to express $D_e/D_0$ as a surface integral over the opening  $\partial \Omega_{in}$ between pores. First, we note that, due to the periodicity of $f$, $\int_{\partial \Omega_{in}} d S_z({\bf r}) f({\bf r})=0$ and the boundary integral in Eq.~(\ref{058423}) can be extended over the whole surface $\partial \Omega$. One can then use the divergence theorem to obtain
\begin{equation}
\frac{D_e}{D_0} = 1- \frac{1}{|\Omega|}\int_{\Omega} d{\bf{r}}\nabla f({\bf r}){\bf e}_z
\end{equation}
Using  the divergence theorem once more, we find
\begin{align}
\frac{D_e}{D_0} = 1+\frac{1}{|\Omega|}\int_{\Omega} d{\bf{r}}z\nabla^2 f({\bf r})-\frac{1}{|\Omega|}\int_{\partial \Omega}dS\ {\bf{n}}\ z\nabla f.
\end{align}
This expression can be simplified by noting that (i) $\nabla^2f=0$ in the bulk, (ii) $ {\bf n}\cdot\nabla f={\bf n }\cdot{\bf e}_z$ at the channel boundary $\partial\Omega_{out}$, and (iii) $\int_{\partial \Omega}dS {\bf{n}}\cdot{\bf e}_z z=\vert\Omega\vert$. Hence,
\begin{equation}
\frac{D_e}{D_0} = \frac{1}{|\Omega|}\int_{\partial \Omega_{in}} d S_z({\bf r}) z - \frac{1}{|\Omega|}\int_{\partial \Omega_{in}} d S z \ {\bf n}\cdot \nabla f. 
\end{equation}
The above integral can now be evaluated using the value of $f$ in the opening between pores, Eqs.~(\ref{7473}),(\ref{0542}), leading to
\begin{align}
\frac{D_e}{D_0} &= \frac{2aL}{|\Omega|} + \frac{\pi L^2}{2|\Omega|} \left[ \frac{1}{\ln (2/\varepsilon)} \right. \nonumber\\
&+ \pi \frac{2G({\bf r}_-,{\bf r}_+) -R({\bf r}_-,{\bf r}_-)-R({\bf r}_+,{\bf r}_+)}{2\ln (2/\varepsilon)^2} \nonumber\\
&\left.+\mathcal{O}\left(\frac{1}{\ln (2/\varepsilon)^3}\right) \right].
\end{align}
Finally under the assumption $V=|\Omega| \gg aL$ we find
\begin{equation}
\label{De_inter}
\frac{D_e}{D_0} \simeq \frac{\pi L^2}{2V} \frac{1}{\ln(2\kappa/\varepsilon)},
\end{equation}
where $\kappa$ is given by $\ln \kappa = (\pi/2) [R({\bf r}_-,{\bf r}_-)+R({\bf r}_+,{\bf r}_+)-2G({\bf r}_-,{\bf r}_+)]$. 

For channels with $H\gg L$, one may use the value of $\kappa$ calculated for a domain that has the shape for the infinite strip $z \in [-L/2,L/2]$ and $r \in [0,\infty[$, for which the Green's function $G$ is given by (see \textit{e.g.} Ref.~\cite{barton1989elements})
\begin{align}
&G({\bf r},{\bf r}_\pm)=-\frac{r}{2L} \nonumber\\
&- \frac{1}{2\pi} \ln \left\{ 4e^{-\frac{\pi r}{L}} \left[ \sinh^2 \frac{\pi r}{2L} + \sin^2 \frac{\pi (z\mp L/2)}{2L} \right] \right\}.
\end{align}
Using this expression we obtain $R({\bf r}_-,{\bf r}_-) =R({\bf r}_+,{\bf r}_+) = - \frac{\ln \pi}{\pi}$ and $G({\bf r}_-,{\bf r}_+) = - \frac{\ln 2}{\pi}$, yielding $\kappa= 2/\pi$. 



\begin{thebibliography}{55}
\expandafter\ifx\csname natexlab\endcsname\relax\def\natexlab#1{#1}\fi
\expandafter\ifx\csname bibnamefont\endcsname\relax
  \def\bibnamefont#1{#1}\fi
\expandafter\ifx\csname bibfnamefont\endcsname\relax
  \def\bibfnamefont#1{#1}\fi
\expandafter\ifx\csname citenamefont\endcsname\relax
  \def\citenamefont#1{#1}\fi
\expandafter\ifx\csname url\endcsname\relax
  \def\url#1{\texttt{#1}}\fi
\expandafter\ifx\csname urlprefix\endcsname\relax\def\urlprefix{URL }\fi
\providecommand{\bibinfo}[2]{#2}
\providecommand{\eprint}[2][]{\url{#2}}

\bibitem[{\citenamefont{Le~Borgne et~al.}(2013)\citenamefont{Le~Borgne, Dentz,
  and Villermaux}}]{leBorgne2013stretching}
\bibinfo{author}{\bibfnamefont{T.}~\bibnamefont{Le~Borgne}},
  \bibinfo{author}{\bibfnamefont{M.}~\bibnamefont{Dentz}}, \bibnamefont{and}
  \bibinfo{author}{\bibfnamefont{E.}~\bibnamefont{Villermaux}},
  \bibinfo{journal}{Phys. Rev. Lett.} \textbf{\bibinfo{volume}{110}},
  \bibinfo{pages}{204501} (\bibinfo{year}{2013}).

\bibitem[{\citenamefont{Dentz et~al.}(2011)\citenamefont{Dentz, Le~Borgne,
  Englert, and Bijeljic}}]{dentz2011mixing}
\bibinfo{author}{\bibfnamefont{M.}~\bibnamefont{Dentz}},
  \bibinfo{author}{\bibfnamefont{T.}~\bibnamefont{Le~Borgne}},
  \bibinfo{author}{\bibfnamefont{A.}~\bibnamefont{Englert}}, \bibnamefont{and}
  \bibinfo{author}{\bibfnamefont{B.}~\bibnamefont{Bijeljic}},
  \bibinfo{journal}{J. Contam. Hydrol.} \textbf{\bibinfo{volume}{120}},
  \bibinfo{pages}{1} (\bibinfo{year}{2011}).

\bibitem[{\citenamefont{Barros et~al.}(2012)\citenamefont{Barros, Dentz, Koch,
  and Nowak}}]{barros2012flow}
\bibinfo{author}{\bibfnamefont{F.~P.} \bibnamefont{Barros}},
  \bibinfo{author}{\bibfnamefont{M.}~\bibnamefont{Dentz}},
  \bibinfo{author}{\bibfnamefont{J.}~\bibnamefont{Koch}}, \bibnamefont{and}
  \bibinfo{author}{\bibfnamefont{W.}~\bibnamefont{Nowak}},
  \bibinfo{journal}{Geophys. Res. Lett.} \textbf{\bibinfo{volume}{39}}
  (\bibinfo{year}{2012}).

\bibitem[{\citenamefont{Bernate and Drazer}(2012)}]{bernate2012stochastic}
\bibinfo{author}{\bibfnamefont{J.~A.} \bibnamefont{Bernate}} \bibnamefont{and}
  \bibinfo{author}{\bibfnamefont{G.}~\bibnamefont{Drazer}},
  \bibinfo{journal}{Phys. Rev. Lett.} \textbf{\bibinfo{volume}{108}},
  \bibinfo{pages}{214501} (\bibinfo{year}{2012}).

\bibitem[{\citenamefont{Brusseau}(1994)}]{brusseau1994transport}
\bibinfo{author}{\bibfnamefont{M.~L.} \bibnamefont{Brusseau}},
  \bibinfo{journal}{Rev. Geophys.} \textbf{\bibinfo{volume}{32}},
  \bibinfo{pages}{285} (\bibinfo{year}{1994}).

\bibitem[{\citenamefont{Condamin et~al.}(2007)\citenamefont{Condamin,
  B{\'e}nichou, Tejedor, Voituriez, and Klafter}}]{Condamin2007}
\bibinfo{author}{\bibfnamefont{S.}~\bibnamefont{Condamin}},
  \bibinfo{author}{\bibfnamefont{O.}~\bibnamefont{B{\'e}nichou}},
  \bibinfo{author}{\bibfnamefont{V.}~\bibnamefont{Tejedor}},
  \bibinfo{author}{\bibfnamefont{R.}~\bibnamefont{Voituriez}},
  \bibnamefont{and} \bibinfo{author}{\bibfnamefont{J.}~\bibnamefont{Klafter}},
  \bibinfo{journal}{Nature} \textbf{\bibinfo{volume}{450}}, \bibinfo{pages}{77}
  (\bibinfo{year}{2007}).

\bibitem[{\citenamefont{Leitmann and Franosch}(2017)}]{leitmann2017time}
\bibinfo{author}{\bibfnamefont{S.}~\bibnamefont{Leitmann}} \bibnamefont{and}
  \bibinfo{author}{\bibfnamefont{T.}~\bibnamefont{Franosch}},
  \bibinfo{journal}{Phys. Rev. Lett.} \textbf{\bibinfo{volume}{118}},
  \bibinfo{pages}{018001} (\bibinfo{year}{2017}).

\bibitem[{\citenamefont{Aminian et~al.}(2016)\citenamefont{Aminian, Bernardi,
  Camassa, Harris, and McLaughlin}}]{aminian2016boundaries}
\bibinfo{author}{\bibfnamefont{M.}~\bibnamefont{Aminian}},
  \bibinfo{author}{\bibfnamefont{F.}~\bibnamefont{Bernardi}},
  \bibinfo{author}{\bibfnamefont{R.}~\bibnamefont{Camassa}},
  \bibinfo{author}{\bibfnamefont{D.~M.} \bibnamefont{Harris}},
  \bibnamefont{and} \bibinfo{author}{\bibfnamefont{R.~M.}
  \bibnamefont{McLaughlin}}, \bibinfo{journal}{Science} p.
  \bibinfo{pages}{0532} (\bibinfo{year}{2016}).

\bibitem[{\citenamefont{Haynes and Vanneste}(2014)}]{haynes2014dispersion1}
\bibinfo{author}{\bibfnamefont{P.}~\bibnamefont{Haynes}} \bibnamefont{and}
  \bibinfo{author}{\bibfnamefont{J.}~\bibnamefont{Vanneste}},
  \bibinfo{journal}{J. Fluid Mech.} \textbf{\bibinfo{volume}{745}},
  \bibinfo{pages}{321} (\bibinfo{year}{2014}).

\bibitem[{\citenamefont{Tzella and Vanneste}(2016)}]{tzella2016dispersion}
\bibinfo{author}{\bibfnamefont{A.}~\bibnamefont{Tzella}} \bibnamefont{and}
  \bibinfo{author}{\bibfnamefont{J.}~\bibnamefont{Vanneste}},
  \bibinfo{journal}{Phys. Rev. Lett.} \textbf{\bibinfo{volume}{117}},
  \bibinfo{pages}{114501} (\bibinfo{year}{2016}).

\bibitem[{\citenamefont{Gu\'erin and Dean}(2015{\natexlab{a}})}]{guerin2015}
\bibinfo{author}{\bibfnamefont{T.}~\bibnamefont{Gu\'erin}} \bibnamefont{and}
  \bibinfo{author}{\bibfnamefont{D.~S.} \bibnamefont{Dean}},
  \bibinfo{journal}{Phys. Rev. Lett.} \textbf{\bibinfo{volume}{115}},
  \bibinfo{pages}{020601} (\bibinfo{year}{2015}{\natexlab{a}}).

\bibitem[{\citenamefont{Brenner}(2013)}]{brenner2013macrotransport}
\bibinfo{author}{\bibfnamefont{H.}~\bibnamefont{Brenner}},
  \emph{\bibinfo{title}{Macrotransport processes}}
  (\bibinfo{publisher}{Elsevier}, \bibinfo{year}{2013}).

\bibitem[{\citenamefont{Taylor}(1953)}]{taylor1953dispersion}
\bibinfo{author}{\bibfnamefont{G.}~\bibnamefont{Taylor}},
  \bibinfo{journal}{Proc. R. Soc. Lon. A} \textbf{\bibinfo{volume}{219}},
  \bibinfo{pages}{186} (\bibinfo{year}{1953}).

\bibitem[{\citenamefont{Van~Kampen}(2007)}]{VanKampen1992}
\bibinfo{author}{\bibfnamefont{N.}~\bibnamefont{Van~Kampen}},
  \emph{\bibinfo{title}{Stochastic Processes in Physics and Chemistry, Third
  Edition}} (\bibinfo{publisher}{North-Holland, Amsterdam},
  \bibinfo{year}{2007}).

\bibitem[{\citenamefont{Dean et~al.}(2007)\citenamefont{Dean, Drummond, and
  Horgan}}]{dean2007effective}
\bibinfo{author}{\bibfnamefont{D.~S.} \bibnamefont{Dean}},
  \bibinfo{author}{\bibfnamefont{I.}~\bibnamefont{Drummond}}, \bibnamefont{and}
  \bibinfo{author}{\bibfnamefont{R.}~\bibnamefont{Horgan}},
  \bibinfo{journal}{J. Stat. Mech. Theory Exp} \textbf{\bibinfo{volume}{2007}},
  \bibinfo{pages}{P07013} (\bibinfo{year}{2007}).

\bibitem[{\citenamefont{Burada et~al.}(2009)\citenamefont{Burada, H{\"a}nggi,
  Marchesoni, Schmid, and Talkner}}]{burada2009diffusion}
\bibinfo{author}{\bibfnamefont{P.~S.} \bibnamefont{Burada}},
  \bibinfo{author}{\bibfnamefont{P.}~\bibnamefont{H{\"a}nggi}},
  \bibinfo{author}{\bibfnamefont{F.}~\bibnamefont{Marchesoni}},
  \bibinfo{author}{\bibfnamefont{G.}~\bibnamefont{Schmid}}, \bibnamefont{and}
  \bibinfo{author}{\bibfnamefont{P.}~\bibnamefont{Talkner}},
  \bibinfo{journal}{ChemPhysChem} \textbf{\bibinfo{volume}{10}},
  \bibinfo{pages}{45} (\bibinfo{year}{2009}).

\bibitem[{\citenamefont{Malgaretti et~al.}(2013)\citenamefont{Malgaretti,
  Pagonabarraga, and Rubi}}]{malgaretti2013entropic}
\bibinfo{author}{\bibfnamefont{P.}~\bibnamefont{Malgaretti}},
  \bibinfo{author}{\bibfnamefont{I.}~\bibnamefont{Pagonabarraga}},
  \bibnamefont{and} \bibinfo{author}{\bibfnamefont{M.}~\bibnamefont{Rubi}},
  \bibinfo{journal}{Frontiers in Physics} \textbf{\bibinfo{volume}{1}},
  \bibinfo{pages}{21} (\bibinfo{year}{2013}).

\bibitem[{\citenamefont{Bressloff and Newby}(2013)}]{bressloff2013stochastic}
\bibinfo{author}{\bibfnamefont{P.~C.} \bibnamefont{Bressloff}}
  \bibnamefont{and} \bibinfo{author}{\bibfnamefont{J.~M.} \bibnamefont{Newby}},
  \bibinfo{journal}{Rev. Mod. Phys.} \textbf{\bibinfo{volume}{85}},
  \bibinfo{pages}{135} (\bibinfo{year}{2013}).

\bibitem[{\citenamefont{Holcman and Schuss}(2013)}]{holcman2013control}
\bibinfo{author}{\bibfnamefont{D.}~\bibnamefont{Holcman}} \bibnamefont{and}
  \bibinfo{author}{\bibfnamefont{Z.}~\bibnamefont{Schuss}},
  \bibinfo{journal}{Reports on Progress in Physics}
  \textbf{\bibinfo{volume}{76}}, \bibinfo{pages}{074601}
  (\bibinfo{year}{2013}).

\bibitem[{\citenamefont{Jacobs}(1967)}]{jac1967}
\bibinfo{author}{\bibfnamefont{M.}~\bibnamefont{Jacobs}},
  \emph{\bibinfo{title}{Diffusion processes}} (\bibinfo{publisher}{Springer,
  New-York}, \bibinfo{year}{1967}).

\bibitem[{\citenamefont{Zwanzig}(1992)}]{zwanzig1992diffusion}
\bibinfo{author}{\bibfnamefont{R.}~\bibnamefont{Zwanzig}}, \bibinfo{journal}{J
  Phys. Chem.} \textbf{\bibinfo{volume}{96}}, \bibinfo{pages}{3926}
  (\bibinfo{year}{1992}).

\bibitem[{\citenamefont{Reguera and Rubi}(2001)}]{reguera2001kinetic}
\bibinfo{author}{\bibfnamefont{D.}~\bibnamefont{Reguera}} \bibnamefont{and}
  \bibinfo{author}{\bibfnamefont{J.}~\bibnamefont{Rubi}},
  \bibinfo{journal}{Phys. Rev. E} \textbf{\bibinfo{volume}{64}},
  \bibinfo{pages}{061106} (\bibinfo{year}{2001}).

\bibitem[{\citenamefont{Kalinay and Percus}(2006)}]{kalinay2006corrections}
\bibinfo{author}{\bibfnamefont{P.}~\bibnamefont{Kalinay}} \bibnamefont{and}
  \bibinfo{author}{\bibfnamefont{J.}~\bibnamefont{Percus}},
  \bibinfo{journal}{Phys. Rev. E} \textbf{\bibinfo{volume}{74}},
  \bibinfo{pages}{041203} (\bibinfo{year}{2006}).

\bibitem[{\citenamefont{Kalinay and
  Percus}(2005{\natexlab{a}})}]{kalinay2005extended}
\bibinfo{author}{\bibfnamefont{P.}~\bibnamefont{Kalinay}} \bibnamefont{and}
  \bibinfo{author}{\bibfnamefont{J.}~\bibnamefont{Percus}},
  \bibinfo{journal}{Phys. Rev. E} \textbf{\bibinfo{volume}{72}},
  \bibinfo{pages}{061203} (\bibinfo{year}{2005}{\natexlab{a}}).

\bibitem[{\citenamefont{Kalinay and
  Percus}(2005{\natexlab{b}})}]{kalinay2005projection}
\bibinfo{author}{\bibfnamefont{P.}~\bibnamefont{Kalinay}} \bibnamefont{and}
  \bibinfo{author}{\bibfnamefont{J.}~\bibnamefont{Percus}},
  \bibinfo{journal}{J. Chem. Phys.} \textbf{\bibinfo{volume}{122}},
  \bibinfo{pages}{204701} (\bibinfo{year}{2005}{\natexlab{b}}).

\bibitem[{\citenamefont{Kalinay and Percus}(2010)}]{kalinay2010mapping}
\bibinfo{author}{\bibfnamefont{P.}~\bibnamefont{Kalinay}} \bibnamefont{and}
  \bibinfo{author}{\bibfnamefont{J.~K.} \bibnamefont{Percus}},
  \bibinfo{journal}{Phys. Rev. E} \textbf{\bibinfo{volume}{82}},
  \bibinfo{pages}{031143} (\bibinfo{year}{2010}).

\bibitem[{\citenamefont{Martens et~al.}(2011)\citenamefont{Martens, Schmid,
  Schimansky-Geier, and H{\"a}nggi}}]{martens2011entropic}
\bibinfo{author}{\bibfnamefont{S.}~\bibnamefont{Martens}},
  \bibinfo{author}{\bibfnamefont{G.}~\bibnamefont{Schmid}},
  \bibinfo{author}{\bibfnamefont{L.}~\bibnamefont{Schimansky-Geier}},
  \bibnamefont{and}
  \bibinfo{author}{\bibfnamefont{P.}~\bibnamefont{H{\"a}nggi}},
  \bibinfo{journal}{Phys. Rev. E} \textbf{\bibinfo{volume}{83}},
  \bibinfo{pages}{051135} (\bibinfo{year}{2011}).

\bibitem[{\citenamefont{Bradley}(2009)}]{bradley2009diffusion}
\bibinfo{author}{\bibfnamefont{R.~M.} \bibnamefont{Bradley}},
  \bibinfo{journal}{Phys. Rev. E} \textbf{\bibinfo{volume}{80}},
  \bibinfo{pages}{061142} (\bibinfo{year}{2009}).

\bibitem[{\citenamefont{Berezhkovskii and Szabo}(2011)}]{berezhkovskii2011time}
\bibinfo{author}{\bibfnamefont{A.}~\bibnamefont{Berezhkovskii}}
  \bibnamefont{and} \bibinfo{author}{\bibfnamefont{A.}~\bibnamefont{Szabo}},
  \bibinfo{journal}{J. Chem. Phys.} \textbf{\bibinfo{volume}{135}},
  \bibinfo{pages}{074108} (\bibinfo{year}{2011}).

\bibitem[{\citenamefont{Dagdug and Pineda}(2012)}]{dagdug2012projection}
\bibinfo{author}{\bibfnamefont{L.}~\bibnamefont{Dagdug}} \bibnamefont{and}
  \bibinfo{author}{\bibfnamefont{I.}~\bibnamefont{Pineda}},
  \bibinfo{journal}{J. Chem. Phys.} \textbf{\bibinfo{volume}{137}},
  \bibinfo{pages}{024107} (\bibinfo{year}{2012}).

\bibitem[{\citenamefont{Valdes and Guzman}(2014)}]{valdes2014fick}
\bibinfo{author}{\bibfnamefont{C.~V.} \bibnamefont{Valdes}} \bibnamefont{and}
  \bibinfo{author}{\bibfnamefont{R.~H.} \bibnamefont{Guzman}},
  \bibinfo{journal}{Phys. Rev. E} \textbf{\bibinfo{volume}{90}},
  \bibinfo{pages}{052141} (\bibinfo{year}{2014}).

\bibitem[{\citenamefont{Lifson and Jackson}(1962)}]{lifson1962self}
\bibinfo{author}{\bibfnamefont{S.}~\bibnamefont{Lifson}} \bibnamefont{and}
  \bibinfo{author}{\bibfnamefont{J.~L.} \bibnamefont{Jackson}},
  \bibinfo{journal}{J. Chem. Phys.} \textbf{\bibinfo{volume}{36}},
  \bibinfo{pages}{2410} (\bibinfo{year}{1962}).

\bibitem[{\citenamefont{Reimann et~al.}(2001)\citenamefont{Reimann, Van~den
  Broeck, Linke, H{\"a}nggi, Rubi, and P{\'e}rez-Madrid}}]{reimann2001giant}
\bibinfo{author}{\bibfnamefont{P.}~\bibnamefont{Reimann}},
  \bibinfo{author}{\bibfnamefont{C.}~\bibnamefont{Van~den Broeck}},
  \bibinfo{author}{\bibfnamefont{H.}~\bibnamefont{Linke}},
  \bibinfo{author}{\bibfnamefont{P.}~\bibnamefont{H{\"a}nggi}},
  \bibinfo{author}{\bibfnamefont{J.~M.} \bibnamefont{Rubi}}, \bibnamefont{and}
  \bibinfo{author}{\bibfnamefont{A.}~\bibnamefont{P{\'e}rez-Madrid}},
  \bibinfo{journal}{Phys. Rev. Lett.} \textbf{\bibinfo{volume}{87}},
  \bibinfo{pages}{010602} (\bibinfo{year}{2001}).

\bibitem[{\citenamefont{Reguera et~al.}(2006)\citenamefont{Reguera, Schmid,
  Burada, Rubi, Reimann, and H{\"a}nggi}}]{reguera2006entropic}
\bibinfo{author}{\bibfnamefont{D.}~\bibnamefont{Reguera}},
  \bibinfo{author}{\bibfnamefont{G.}~\bibnamefont{Schmid}},
  \bibinfo{author}{\bibfnamefont{P.~S.} \bibnamefont{Burada}},
  \bibinfo{author}{\bibfnamefont{J.~M.} \bibnamefont{Rubi}},
  \bibinfo{author}{\bibfnamefont{P.}~\bibnamefont{Reimann}}, \bibnamefont{and}
  \bibinfo{author}{\bibfnamefont{P.}~\bibnamefont{H{\"a}nggi}},
  \bibinfo{journal}{Phys. Rev. Lett.} \textbf{\bibinfo{volume}{96}},
  \bibinfo{pages}{130603} (\bibinfo{year}{2006}).

\bibitem[{\citenamefont{Berezhkovskii et~al.}(2009)\citenamefont{Berezhkovskii,
  Barzykin, and Zitserman}}]{berezhkovskii2009one}
\bibinfo{author}{\bibfnamefont{A.~M.} \bibnamefont{Berezhkovskii}},
  \bibinfo{author}{\bibfnamefont{A.~V.} \bibnamefont{Barzykin}},
  \bibnamefont{and} \bibinfo{author}{\bibfnamefont{V.~Y.}
  \bibnamefont{Zitserman}}, \bibinfo{journal}{J. Chem. Phys.}
  \textbf{\bibinfo{volume}{131}}, \bibinfo{pages}{224110}
  (\bibinfo{year}{2009}).

\bibitem[{\citenamefont{Antipov et~al.}(2013)\citenamefont{Antipov, Barzykin,
  Berezhkovskii, Makhnovskii, Zitserman, and Aldoshin}}]{antipov2013effective}
\bibinfo{author}{\bibfnamefont{A.~E.} \bibnamefont{Antipov}},
  \bibinfo{author}{\bibfnamefont{A.~V.} \bibnamefont{Barzykin}},
  \bibinfo{author}{\bibfnamefont{A.~M.} \bibnamefont{Berezhkovskii}},
  \bibinfo{author}{\bibfnamefont{Y.~A.} \bibnamefont{Makhnovskii}},
  \bibinfo{author}{\bibfnamefont{V.~Y.} \bibnamefont{Zitserman}},
  \bibnamefont{and} \bibinfo{author}{\bibfnamefont{S.~M.}
  \bibnamefont{Aldoshin}}, \bibinfo{journal}{Phys. Rev. E}
  \textbf{\bibinfo{volume}{88}}, \bibinfo{pages}{054101}
  (\bibinfo{year}{2013}).

\bibitem[{\citenamefont{Bosi et~al.}(2012)\citenamefont{Bosi, Ghosh, and
  Marchesoni}}]{bosi2012analytical}
\bibinfo{author}{\bibfnamefont{L.}~\bibnamefont{Bosi}},
  \bibinfo{author}{\bibfnamefont{P.~K.} \bibnamefont{Ghosh}}, \bibnamefont{and}
  \bibinfo{author}{\bibfnamefont{F.}~\bibnamefont{Marchesoni}},
  \bibinfo{journal}{J. Chem. Phys.} \textbf{\bibinfo{volume}{137}},
  \bibinfo{pages}{174110} (\bibinfo{year}{2012}).

\bibitem[{\citenamefont{Borromeo and Marchesoni}(2010)}]{borromeo2010particle}
\bibinfo{author}{\bibfnamefont{M.}~\bibnamefont{Borromeo}} \bibnamefont{and}
  \bibinfo{author}{\bibfnamefont{F.}~\bibnamefont{Marchesoni}},
  \bibinfo{journal}{Chem. Phys.} \textbf{\bibinfo{volume}{375}},
  \bibinfo{pages}{536} (\bibinfo{year}{2010}).

\bibitem[{\citenamefont{Marchesoni}(2010)}]{marchesoni2010mobility}
\bibinfo{author}{\bibfnamefont{F.}~\bibnamefont{Marchesoni}},
  \bibinfo{journal}{J. Chem. Phys.} \textbf{\bibinfo{volume}{132}},
  \bibinfo{pages}{166101} (\bibinfo{year}{2010}).

\bibitem[{\citenamefont{Pineda et~al.}(2011)\citenamefont{Pineda, Vazquez,
  Berezhkovskii, and Dagdug}}]{pineda2011diffusion}
\bibinfo{author}{\bibfnamefont{I.}~\bibnamefont{Pineda}},
  \bibinfo{author}{\bibfnamefont{M.-V.} \bibnamefont{Vazquez}},
  \bibinfo{author}{\bibfnamefont{A.~M.} \bibnamefont{Berezhkovskii}},
  \bibnamefont{and} \bibinfo{author}{\bibfnamefont{L.}~\bibnamefont{Dagdug}},
  \bibinfo{journal}{J. Chem. Phys.} \textbf{\bibinfo{volume}{135}},
  \bibinfo{pages}{224101} (\bibinfo{year}{2011}).

\bibitem[{\citenamefont{Berezhkovskii et~al.}(2003)\citenamefont{Berezhkovskii,
  Zitserman, and Shvartsman}}]{berezhkovskii2003diffusivity}
\bibinfo{author}{\bibfnamefont{A.~M.} \bibnamefont{Berezhkovskii}},
  \bibinfo{author}{\bibfnamefont{V.~Y.} \bibnamefont{Zitserman}},
  \bibnamefont{and} \bibinfo{author}{\bibfnamefont{S.~Y.}
  \bibnamefont{Shvartsman}}, \bibinfo{journal}{J. Chem. Phys.}
  \textbf{\bibinfo{volume}{118}}, \bibinfo{pages}{7146} (\bibinfo{year}{2003}).

\bibitem[{\citenamefont{Gu\'erin and
  Dean}(2015{\natexlab{b}})}]{Guerin2015Kubo}
\bibinfo{author}{\bibfnamefont{T.}~\bibnamefont{Gu\'erin}} \bibnamefont{and}
  \bibinfo{author}{\bibfnamefont{D.~S.} \bibnamefont{Dean}},
  \bibinfo{journal}{Phys. Rev. E} \textbf{\bibinfo{volume}{92}},
  \bibinfo{pages}{062103} (\bibinfo{year}{2015}{\natexlab{b}}).

\bibitem[{\citenamefont{Dorfman and Yariv}(2014)}]{dorfman2014assessing}
\bibinfo{author}{\bibfnamefont{K.~D.} \bibnamefont{Dorfman}} \bibnamefont{and}
  \bibinfo{author}{\bibfnamefont{E.}~\bibnamefont{Yariv}}, \bibinfo{journal}{J.
  Chem. Phys.} \textbf{\bibinfo{volume}{141}}, \bibinfo{pages}{044118}
  (\bibinfo{year}{2014}).

\bibitem[{\citenamefont{Berezhkovskii et~al.}(2015)\citenamefont{Berezhkovskii,
  Dagdug, and Bezrukov}}]{berezhkovskii2015biased}
\bibinfo{author}{\bibfnamefont{A.~M.} \bibnamefont{Berezhkovskii}},
  \bibinfo{author}{\bibfnamefont{L.}~\bibnamefont{Dagdug}}, \bibnamefont{and}
  \bibinfo{author}{\bibfnamefont{S.~M.} \bibnamefont{Bezrukov}},
  \bibinfo{journal}{J. Chem. Phys.} \textbf{\bibinfo{volume}{142}},
  \bibinfo{pages}{134101} (\bibinfo{year}{2015}).

\bibitem[{\citenamefont{Dean and Jansons}(1993)}]{dean1993brownian}
\bibinfo{author}{\bibfnamefont{D.~S.} \bibnamefont{Dean}} \bibnamefont{and}
  \bibinfo{author}{\bibfnamefont{K.~M.} \bibnamefont{Jansons}},
  \bibinfo{journal}{J. Stat. Phys.} \textbf{\bibinfo{volume}{70}},
  \bibinfo{pages}{1313} (\bibinfo{year}{1993}).

\bibitem[{\citenamefont{Dagdug et~al.}(2007)\citenamefont{Dagdug,
  Berezhkovskii, Makhnovskii, and Zitserman}}]{dagdug2007transient}
\bibinfo{author}{\bibfnamefont{L.}~\bibnamefont{Dagdug}},
  \bibinfo{author}{\bibfnamefont{A.~M.} \bibnamefont{Berezhkovskii}},
  \bibinfo{author}{\bibfnamefont{Y.~A.} \bibnamefont{Makhnovskii}},
  \bibnamefont{and} \bibinfo{author}{\bibfnamefont{V.~Y.}
  \bibnamefont{Zitserman}}, \bibinfo{journal}{J. Chem. Phys.}
  \textbf{\bibinfo{volume}{127}}, \bibinfo{pages}{224712}
  (\bibinfo{year}{2007}).

\bibitem[{\citenamefont{Dagdug et~al.}(2012)\citenamefont{Dagdug, Vazquez,
  Berezhkovskii, Zitserman, and Bezrukov}}]{dagdug2012diffusion}
\bibinfo{author}{\bibfnamefont{L.}~\bibnamefont{Dagdug}},
  \bibinfo{author}{\bibfnamefont{M.-V.} \bibnamefont{Vazquez}},
  \bibinfo{author}{\bibfnamefont{A.~M.} \bibnamefont{Berezhkovskii}},
  \bibinfo{author}{\bibfnamefont{V.~Y.} \bibnamefont{Zitserman}},
  \bibnamefont{and} \bibinfo{author}{\bibfnamefont{S.~M.}
  \bibnamefont{Bezrukov}}, \bibinfo{journal}{J. Chem. Phys.}
  \textbf{\bibinfo{volume}{136}}, \bibinfo{pages}{204106}
  (\bibinfo{year}{2012}).

\bibitem[{\citenamefont{Crank}(1979)}]{crank1979mathematics}
\bibinfo{author}{\bibfnamefont{J.}~\bibnamefont{Crank}},
  \emph{\bibinfo{title}{The mathematics of diffusion}}
  (\bibinfo{publisher}{Oxford university press}, \bibinfo{year}{1979}).

\bibitem[{\citenamefont{Holcman et~al.}(2011)\citenamefont{Holcman, Hoze, and
  Schuss}}]{holcman2011narrow}
\bibinfo{author}{\bibfnamefont{D.}~\bibnamefont{Holcman}},
  \bibinfo{author}{\bibfnamefont{N.}~\bibnamefont{Hoze}}, \bibnamefont{and}
  \bibinfo{author}{\bibfnamefont{Z.}~\bibnamefont{Schuss}},
  \bibinfo{journal}{Phys. Rev. E} \textbf{\bibinfo{volume}{84}},
  \bibinfo{pages}{021906} (\bibinfo{year}{2011}).

\bibitem[{\citenamefont{Holcman and Schuss}(2012)}]{holcman2012brownian}
\bibinfo{author}{\bibfnamefont{D.}~\bibnamefont{Holcman}} \bibnamefont{and}
  \bibinfo{author}{\bibfnamefont{Z.}~\bibnamefont{Schuss}},
  \bibinfo{journal}{Multiscale Modeling \& Simulation}
  \textbf{\bibinfo{volume}{10}}, \bibinfo{pages}{1204} (\bibinfo{year}{2012}).

\bibitem[{\citenamefont{Guerrier and Holcman}(2014)}]{guerrier2014brownian}
\bibinfo{author}{\bibfnamefont{C.}~\bibnamefont{Guerrier}} \bibnamefont{and}
  \bibinfo{author}{\bibfnamefont{D.}~\bibnamefont{Holcman}},
  \bibinfo{journal}{Eur. Phys. J Special Topics}
  \textbf{\bibinfo{volume}{223}}, \bibinfo{pages}{3273} (\bibinfo{year}{2014}).

\bibitem[{\citenamefont{Ward and Keller}(1993)}]{ward1993}
\bibinfo{author}{\bibfnamefont{M.~J.} \bibnamefont{Ward}} \bibnamefont{and}
  \bibinfo{author}{\bibfnamefont{J.~B.} \bibnamefont{Keller}},
  \bibinfo{journal}{SIA M J. Appl Math} \textbf{\bibinfo{volume}{53}},
  \bibinfo{pages}{770} (\bibinfo{year}{1993}).

\bibitem[{\citenamefont{Pillay et~al.}(2010)\citenamefont{Pillay, Ward, Peirce,
  and Kolokolnikov}}]{pillay2010asymptotic}
\bibinfo{author}{\bibfnamefont{S.}~\bibnamefont{Pillay}},
  \bibinfo{author}{\bibfnamefont{M.~J.} \bibnamefont{Ward}},
  \bibinfo{author}{\bibfnamefont{A.}~\bibnamefont{Peirce}}, \bibnamefont{and}
  \bibinfo{author}{\bibfnamefont{T.}~\bibnamefont{Kolokolnikov}},
  \bibinfo{journal}{Multiscale Modeling \& Simulation}
  \textbf{\bibinfo{volume}{8}}, \bibinfo{pages}{803} (\bibinfo{year}{2010}).

\bibitem[{\citenamefont{Barton}(1989)}]{barton1989elements}
\bibinfo{author}{\bibfnamefont{G.}~\bibnamefont{Barton}},
  \emph{\bibinfo{title}{Elements of Green's functions and propagation}}
  (\bibinfo{publisher}{Clarendon Press, Oxford}, \bibinfo{year}{1989}).

\bibitem[{\citenamefont{B{\'e}nichou and Voituriez}(2008)}]{Benichou2008}
\bibinfo{author}{\bibfnamefont{O.}~\bibnamefont{B{\'e}nichou}}
  \bibnamefont{and}
  \bibinfo{author}{\bibfnamefont{R.}~\bibnamefont{Voituriez}},
  \bibinfo{journal}{Phys Rev Lett} \textbf{\bibinfo{volume}{100}},
  \bibinfo{pages}{168105} (\bibinfo{year}{2008}).

\end{thebibliography}

\end{document}